\documentclass{aa}

\usepackage{comment}
\usepackage{graphicx}
\usepackage{txfonts}
\usepackage{lipsum}
\usepackage{subcaption} 
                               
\usepackage{lscape} 
\usepackage{placeins} 
\usepackage{booktabs}
\usepackage{threeparttable}
\usepackage{tabularx}
\usepackage{makecell} 
 \usepackage{tikz}
 \usetikzlibrary{arrows.meta, calc, decorations.pathreplacing}
\usepackage{cuted}
\usepackage{capt-of}
\usepackage[hidelinks]{hyperref}

\begin{document}

   \title{4U~1624$-$490 as a possible intermediate-mass X-ray binary: constraints from XRISM and NuSTAR spectroscopy}

   \author{R. Iaria\inst{1}\fnmsep\thanks{Corresponding author: rosario.iaria@unipa.it},
   A. Anitra \inst{1}, T. Di Salvo\inst{1},  A. Sanna\inst{2,3}, F. Barra\inst{4,1},  L. Burderi\inst{2}, C. Maraventano\inst{1,5}, S. Caserta\inst{1,4}, W. Leone\inst{2}, C. Miceli\inst{4,6}}

 \titlerunning{4U 1624$-$490 as a possible intermediate-mass X-ray binary}
\authorrunning{R. Iaria et al.}

\institute{Università degli Studi di Palermo, Dipartimento di Fisica e Chimica, Via Archirafi 36, 90123 Palermo, Italy;   \and Dipartimento di Fisica, Universit\`a degli Studi di Cagliari, SP Monserrato-Sestu, KM 0.7, Monserrato, 09042, Italy;    \and Osservatorio Astronomico di Cagliari, Istituto nazionale di astrofisica, via della Scienza 5, Selargius (CA), I-09047, Italia; \and INAF -- IASF Palermo, Via U. La Malfa 153, 90146 Palermo, Italy; \and INAF -- Osservatorio Astronomica di Brera, Via E. Bianchi 46, I-23807, Merate (LC), Italy;  \and Departament de Física, EEBE, Universitat Politècnica de Catalunya, Av. Eduard Maristany 16, 08019, Barcelona, Spain
}

   \date{}

\abstract
{High-inclination dipping neutron-star X-ray binaries provide a unique
view of accretion-flow structure because the line of sight intercepts
the outer-disc absorber and ionised plasma above the disc. Their orbital
variability is generally dominated by absorption, although disc
reflection may also contribute.}
{We investigate the orbital evolution of the ionised absorber and
reflection component in 4U~1624$-$490 to constrain the kinematics and
nature of its highly ionised plasma.}
{We analysed two XRISM/Resolve observations with simultaneous
NuSTAR coverage. The data were divided into seven orbital-phase
intervals covering persistent and dipping states. We performed
phase-resolved fits including neutral partial covering, a photoionised
absorber, thermal Comptonisation, and ionised disc reflection.}
{Fe\,{\sc xxv} and Fe\,{\sc xxvi} absorption features are detected over
most of the orbit. The absorber column density, ionisation state,
covering fraction, and projected velocity vary with orbital phase in agreement with previous results. It is
systematically blueshifted, with line-of-sight velocities of approximately
$85$--$440\ \mathrm{km\,s^{-1}}$, consistent with a structured outflow or
vertically extended plasma. Absorption-only models leave broad Fe--K
residuals, whereas ionised reflection systematically improves the fits
and varies in visibility along the orbit. The velocity modulation has a
semi-amplitude $K=210\pm14\ \mathrm{km\,s^{-1}}$ and, under an orbital
interpretation, implies a mass function
$f(M)=0.83\pm0.17\,M_\odot$.}
{The orbital spectral variability therefore requires the combined effects
of ionised absorption, Comptonised emission, and disc reflection. Residual
structure in the blue wing of Fe\,{\sc xxvi} may indicate additional
kinematic complexity close to the dip; nevertheless, the velocity modulation
and its fitted parameters remain consistent when the two dip-adjacent
measurements are excluded. If this modulation traces the neutron-star orbital
motion, the inferred donor mass, $M_2=2.66\pm0.32\,M_\odot$, is compatible
with a Roche-lobe-filling B9~V--B9.5~V star. Under this interpretation,
4U~1624$-$490 may be an intermediate-mass rather than a classical low-mass
X-ray binary}


   \maketitle
\nolinenumbers

\section{Introduction}

High-inclination low-mass X-ray binaries (LMXBs) offer one of the best observational probes of the vertical structure of accretion flows. In these systems, the line of sight intercepts the outer accretion disc atmosphere and the stream--disc impact region, often producing recurrent X-ray dips on the orbital period. Because these dips are strongly energy dependent, they have long been interpreted as the signature of structured absorbing material at the disc rim, while the residual flux observed during deep obscuration has been associated with scattering and reprocessing in an accretion-disc corona or extended atmosphere \citep{WhiteHolt1982, Frank1987, ChurchBalucinska2004}. High-inclination neutron-star binaries therefore provide a privileged laboratory for investigating accretion geometry, disc atmospheres, and circumsource absorption.

High-resolution X-ray spectroscopy has substantially refined this picture. In particular, the detection of narrow absorption lines from highly ionised species, especially Fe\,{\sc xxv} and Fe\,{\sc xxvi}, has shown that the spectral evolution of dipping LMXBs cannot be understood solely in terms of neutral obscuration by cold material at the outer disc edge. Instead, a photoionised plasma is now recognised as a key ingredient of these systems, and in many dippers the observed orbital variability can be described primarily by changes in the column density, ionisation state, and covering fraction of the ionised absorber \citep{Boirin2005, DiazTrigo2006, DiazTrigo2009, Iaria2007b}. More generally, the strong association between ionised absorbers and high-inclination systems suggests a close connection with disc atmospheres and winds, making dipping LMXBs ideal targets in which to investigate the transition between static atmospheres, failed winds, and genuine outflows \citep{Ponti2012}.

Within this class, 4U~1624$-$490 is one of the most remarkable systems. Known as the ``Big Dipper'', it is a persistently bright neutron-star X-ray binary with a long orbital period of about 21~hr and broad, complex dipping activity \citep{Watson1985, Smale2001}. Early RXTE observations showed that dipping remains detectable up to $\sim15$~keV, already indicating that simple soft-X-ray photoelectric absorption is insufficient to explain the full phenomenology and favouring instead a picture in which the dips involve progressive obscuration of an extended Comptonising region \citep{Smale2001, ChurchBalucinska2004}.

Subsequent Chandra and XMM-Newton observations demonstrated that 4U~1624$-$490 hosts a rich and complex absorbing environment. From the analysis of the X-ray dust-scattering halo, \citet{Xiang_07} estimated an interstellar column density of $N_{\rm H}^{\rm sca}\sim(3.6$--$4.8)\times10^{22}\ {\rm cm^{-2}}$, significantly lower than the total absorption inferred from broadband X-ray spectroscopy, thus implying that a substantial fraction of the obscuration is local to the binary. 
Consistently, \citet{Xiang_09} found evidence for an additional local neutral absorber. The presence of highly ionised absorption in 4U~1624--490 was first reported by \citet{Parmar_02}, who detected Fe,{\sc xxv} and Fe,{\sc xxvi} absorption lines. Subsequent high-resolution spectroscopy confirmed these features and characterised the absorbing medium as dense, photoionised plasma \citep{Iaria2007b}. More recent time- and flux-resolved analyses further showed that the dip evolution is governed by a clumpy, multiphase absorber, including both ionised and colder components, reinforcing the view that the outer accretion flow is highly inhomogeneous \citep{Caruso_26}.

At the same time, broadband spectroscopy and X-ray polarimetry have shown that absorption alone is not sufficient to describe the source. Recent IXPE-based studies indicate that the observed polarisation properties of 4U~1624$-$490 are difficult to explain with Comptonisation alone and instead require an additional contribution from disc reflection and/or an extended slab-like corona \citep{Saade2024, Gnarini_24}.
These results suggest that reprocessed and/or scattered radiation is an
important ingredient of the high-inclination accretion geometry, although
polarimetry alone does not uniquely separate disc reflection from wind
scattering or re-emission \citep[see e.g.][]{Nitindala_25}. 

Despite this progress, two major questions remain open. First, the physical nature of the highly ionised plasma is still uncertain: the observed absorption may arise in a static disc atmosphere, in a thermally inflated but only partially escaping structure, or in a genuine accretion-disc wind \citep{Diaz2016, Begelman1983, Woods1996, Done2018}. Second, the orbital interplay between absorber kinematics and reflection has not yet been mapped with the spectral resolution required to cleanly resolve the Fe--K complex in a bright dipping neutron-star binary. This is precisely the regime in which XRISM is transformative. Thanks to the non-dispersive microcalorimeter spectroscopy provided by Resolve, XRISM enables accurate measurements of line centroids, widths, and blends in the Fe--K band, opening a new window on the kinematics and physical conditions of photoionised plasmas in compact binaries \citep{Tashiro2025, Ishisaki2025}.

A recent joint XRISM+NuSTAR analysis of the same dataset by
\cite{DiazTrigo2026}  revealed a nearly sinusoidal orbital
modulation of the velocity shift of the highly ionised absorber.
Together with the persistent blueshift of the lines, this behaviour
was interpreted as evidence for an outflowing wind. The inferred
dynamical parameters, however, depended on the treatment of the
complex absorption close to the dip.

\cite{DiazTrigo2026}  obtained three estimates of the
velocity semi-amplitude using different treatments of the
absorption spectrum. Fits to the Fe K$\beta$ and Ni absorption
lines in the 7.5--9~keV band yielded
$K=229\pm14$~km~s$^{-1}$, while their single-absorber
photoionisation model including re-emission (Model~2c)
gave $K=200\pm6$~km~s$^{-1}$. Introducing a second absorber
in the post-dip interval p0 yielded
$K=177\pm9$~km~s$^{-1}$. 
The corresponding mass functions reported by those authors
are 1.08, 0.71, and $0.50\,M_\odot$, respectively. They
favoured the two-absorber interpretation as more compatible
with a low-mass companion, while explicitly acknowledging
the possibility of an intermediate-mass donor when a single
absorber is considered.

In this work, we independently reanalyse the same phase-resolved
XRISM/Resolve and simultaneous NuSTAR observations with two
complementary aims. First, we exploit the broadband coverage provided
by NuSTAR to test explicitly for an ionised disc-reflection
component and to determine whether the broadband spectral
decomposition affects the centroids of the narrow Fe--K absorption
lines. Second, we revisit the orbital modulation of the absorber
velocity using a uniform spectral description across orbital phase.
This analysis yields a revised velocity semi-amplitude and allows us
to examine how the different treatment of the complex dip-adjacent
spectra affects the inferred mass function and, conditionally, the
mass and evolutionary nature of the donor star.

This paper is organised as follows. In Sect.~\ref{sec:obs} we describe the XRISM and NuSTAR observations and the data reduction procedure. In Sect.~\ref{sec:analysis} we present the phase-resolved spectral analysis, and in Sect.~\ref{sec:discussion} we discuss the implications of our results for the accretion geometry, absorber dynamics, and evolutionary nature of the system.

\begin{table}
\caption{Observation log and orbital-phase selection for the XRISM/Resolve and simultaneous NuSTAR observations.}
\label{tab:obslog_phase}
\centering
\scriptsize
\setlength{\tabcolsep}{2pt}
\renewcommand{\arraystretch}{0.95}

\begin{tabular}{@{}llcc@{}}
\hline\hline
\multicolumn{4}{c}{A) Observation log}\\
\hline
Instr. & ObsID & UTC interval & Exp. \\
       & & & (ks) \\
\hline
\multicolumn{4}{c}{Obs.\ 1}\\
\hline
Resolve & 300040010 & 29 Mar 23:56 -- 31 Mar 02:02 & 67.8 \\
NuSTAR & 30902018002 & 30 Mar 00:04 -- 30 Mar 11:46 & 18.7 \\
\hline
\multicolumn{4}{c}{Obs.\ 2}\\
\hline
Resolve & 300040020 & 04 Apr 05:06 -- 05 Apr 06:44 & 57.2 \\
NuSTAR & 30902018004 & 04 Apr 16:21 -- 05 Apr 02:53 & 18.9 \\
\hline
\multicolumn{4}{c}{B) Orbital-phase selection}\\
\hline
Int. & Phase & \multicolumn{2}{c}{Exposure (ks)}\\
\cline{3-4}
 & & Resolve & NuSTAR \\
\hline
\multicolumn{4}{c}{Obs.\ 1}\\
\hline
1 & 0.95--1.12 & 12.4 & 5.4 \\
2 & 0.12--0.25 & 13.4 & 4.1 \\
3 & 0.25--0.32 & 3.9 & 2.3 \\
4 & 0.32--0.47 & 9.3 & 5.4 \\
5 & 0.47--0.71 & 14.7 & 1.5 \\
6 & 0.71--0.87 & 8.8 & -- \\
7 & 0.87--0.95 & 4.8 & -- \\
\hline
\multicolumn{4}{c}{Obs.\ 2}\\
\hline
1 & 0.95--1.12 & 20.0 & 2.1 \\
2 & 0.12--0.25 & 9.3 & -- \\
3 & 0.25--0.32 & 3.0 & -- \\
4 & 0.32--0.47 & 0.8 & -- \\
5 & 0.47--0.71 & 9.5 & 8.5 \\
6 & 0.71--0.87 & 9.4 & 5.5 \\
7 & 0.87--0.95 & 4.7 & 2.6 \\
\hline\hline
\end{tabular}

\tablefoot{ Panel A reports the observation identifiers, UTC intervals, and net screened exposures for Obs.~1 and Obs.~2. Panel B lists the orbital-phase intervals adopted for the phase-resolved analysis together with the corresponding screened exposures for Resolve and NuSTAR. The orbital phases were computed using the ephemeris of \citet{Liao_15}, with $\phi=0$ defined at the dip centre.
}

\end{table}

\section{Observations and data reduction}
\label{sec:obs}

The X-ray Imaging and Spectroscopy Mission (XRISM; \citealt{Tashiro2025}) observed 4U~1624$-$490 twice in 2024
(ObsIDs 300040010 and   300040020), with two $\sim$90~ks pointings separated by four days (see Table~\ref{tab:obslog_phase}). Each observation individually covers more than one orbital period. During both pointings, the Resolve microcalorimeter \citep{Ishisaki_25} was operated with the open filter.

The XRISM/Resolve data were reprocessed and analysed using the mission-specific XRISM FTOOLS distributed within HEASoft~v6.35.2, together with the XRISM CALDB~v11 released in March 2025.
Following the recommendations of the XRISM ABC Guide (v3.1)\footnote{\url{https://heasarc.gsfc.nasa.gov/docs/xrism/analysis/abc_guide/xrism_abc.pdf}}, we extracted Resolve spectra using the high-resolution primary (Hp) events only, excluded pixel~27 because of its irregular gain behaviour, and verified on the unfiltered event file that the Pulse Shape Processor (PSP) count rate remained below the nominal limit of 50~counts~s$^{-1}$ per quadrant throughout the intervals retained for the scientific analysis; since our spectral fitting is restricted to energies above 2~keV, the additional $PI \geq 600$ selection associated with the pulse-height versus rise-time screening does not affect the fitted energy range.

Following the current XRISM/Resolve analysis recommendations, we inspected the population of low-resolution secondary (Ls) events in the cleaned event files. In the 2--10~keV band, the Ls-event fraction is approximately 3.1\% on average, ranging from approximately 2.4\% to 3.8\% in narrower energy intervals, with the largest contribution occurring in the brightest pixels. All scientific spectra were extracted using high-resolution primary (Hp) events only. However, because the effective area appropriate for an Hp-only spectrum depends on the event-grade branching ratios adopted during response generation, excluding Ls events from the fitted spectrum does not by itself remove the associated systematic uncertainty on the absolute flux.

We therefore generated two limiting response files following the XRISM recommendations. The first was calculated with \texttt{includels=yes}, assuming that all Ls events originate from the source, while the second was calculated with \texttt{includels=no}, assuming that all Ls events are non-X-ray events. These two cases bracket, respectively, the lower and upper limits of the effective area and hence the upper and lower limits of the inferred absolute flux. The same Hp-only spectrum was fitted with both responses, with all physical model parameters tied and only a multiplicative normalisation constant allowed to vary. Fixing the constant associated with the \texttt{includels=no} response to unity, we obtained a value of approximately 1.03 for the \texttt{includels=yes} response. We therefore estimate that the uncertainty associated with the Resolve event-grade branching ratios introduces an approximately 3\% systematic range in the absolute flux. This difference is derived directly from the comparison between the two limiting responses and is not inferred from the globally integrated Ls-event fraction. The spectral shape and the narrow absorption-line parameters remained unchanged within their statistical uncertainties.

After standard screening, the net Resolve exposures used in this work are 67.8 ks for ObsID 300040010 (Obs.~1) and 57.2 ks for ObsID 300040020 (Obs.~2). The corresponding light curves are shown in Fig.~\ref{fig:curves}. Both observations exhibit the same overall phenomenology, namely, broad dipping intervals separated by long persistent phases, while sampling complementary portions of the orbital cycle.  

The Nuclear Spectroscopic Telescope Array \citep[NuSTAR][]{Harrison2013}  observed 4U~1624$-$490 simultaneously with both XRISM pointings (ObsIDs 30902018002 and 30902018004). Source and background products for FPMA and FPMB were generated with the standard \textsc{nupipeline} and \textsc{nuproducts} tasks, adopting a circular extraction region of radius 120 arcsec for the source and a source-free circular region for the background. After standard screening and cleaning, the net exposures used in the analysis are 18.7 ks for Obs.~1 and 18.9 ks for Obs.~2.

\subsection{Barycentric correction and orbital-phase assignment}
\label{subsec:bary_phase}

\begin{figure}
        \centering
        \includegraphics[width=1\hsize]{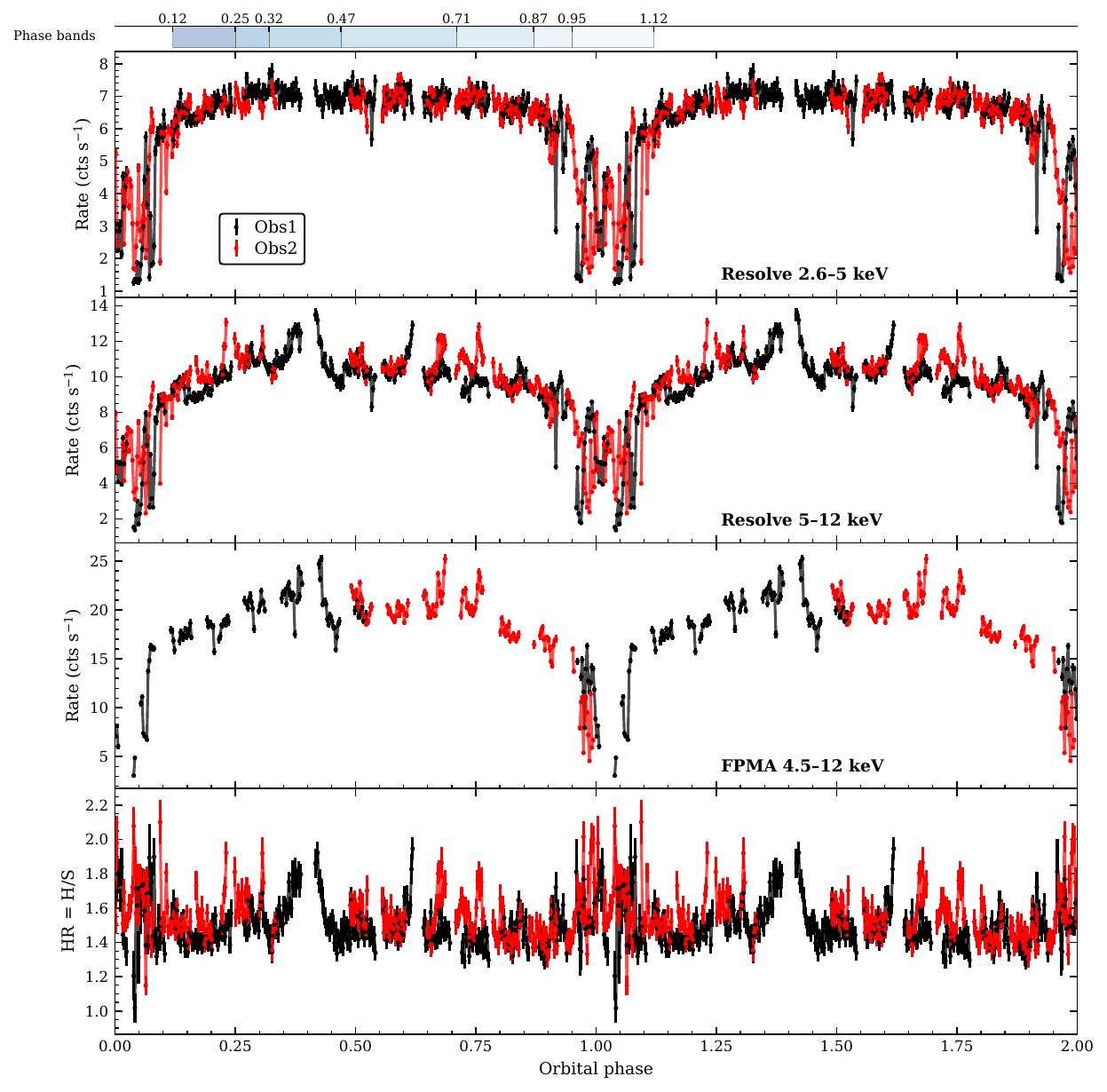}
\caption{Orbital-phase-folded light curves of the two XRISM/Resolve observations and the simultaneous NuSTAR/FPMA data, rebinned into 400 phase bins and repeated over two orbital cycles for clarity. Obs.~1 and Obs.~2 are shown in black and red, respectively. The vertical markers at the top indicate the orbital-phase intervals adopted for the phase-resolved spectral analysis.}
        \label{fig:folded_curves}
    \end{figure}

All event times were converted to the Solar System barycentre using the XRISM orbit file and the task \texttt{barycorr}, adopting the Barycentric Dynamical Time (TDB) as the reference timescale. An analogous procedure was applied to the simultaneous NuSTAR event files in order to generate barycentre-corrected   FPMA and FPMB light curves on the same time system. We then computed the orbital phase according to
\[
T_n\,({\rm MJD,TDB}) = 50088.1618(57) + N \times 0.869896(1),
\]
that is, using $T_0 = 50088.1618(57)$~MJD (TDB) and $P_{\rm orb}=0.869896(1)$~d, and defining $\phi=0$ at the dip centre \citep{Liao_15}.

When folded on the orbital period, the light curves of Obs.~1 and Obs.~2 show a consistent morphology in all bands (Fig.~\ref{fig:folded_curves}). The dip is centred around $\phi \simeq 0$, followed by a rapid recovery and a long persistent interval extending to $\phi \sim 0.8$--0.9 before the onset of the next dip. The dip is deeper in the soft band (2.6--5 keV; Fig.~\ref{fig:folded_curves}, top panel), while the hardness ratio (HR), defined as the ratio between the count rate in the hard band (5--12 keV) and that in the soft band, increases during the low-flux phases, indicating spectral hardening during dipping (Fig.~\ref{fig:folded_curves}, bottom panel). The jagged dip profile may reflect a clumpy ionised absorber along the line of sight \citep[see][and references therein]{Caruso_26}.  
The folded NuSTAR profiles are consistent with the Resolve behaviour, confirming that the orbital modulation remains significant above 4.5 keV as well.

The phase intervals adopted for the phase-resolved analysis, namely 0.12--0.25, 0.25--0.32, 0.32--0.47, 0.47--0.71, 0.71--0.87, 0.87--0.95, and 0.95--1.12, were chosen to track the main morphological changes of the folded orbital profile. Phase-resolved good time intervals (GTIs) were then created for all instruments. The resulting screened exposures are listed in Table~\ref{tab:obslog_phase}. The Resolve data provide almost complete coverage of the orbital cycle in both observations, whereas the simultaneous NuSTAR coverage is more discontinuous because of the shorter net exposure and observational gaps.

We extracted Resolve spectra from the cleaned event files using \texttt{XSELECT}, applying the same grade and pixel
selections adopted in Sect.~\ref{sec:obs}.
Redistribution matrices (RMFs) were generated with \texttt{rslmkrmf} and ancillary response files (ARFs) with
\texttt{xaarfgen}, assuming a point source at the nominal aim point unless otherwise stated.
All response products were produced consistently for each phase-resolved spectrum (and for each dip selection, when used),
ensuring that the pixel list and event grades match between spectra and responses.
Because the instrumental non-X-ray background cannot be directly estimated from a local off-source region for Resolve,
we generated background spectra with \texttt{rslnxbgen}, using the appropriate night-Earth database and matching the same
pixel/grade selections for the source spectra.
Source and background products, therefore, share consistent scaling keywords and selections.

Finally, each spectrum was grouped with the FTOOL \texttt{ftgrouppha} to ensure at least 25 counts in each energy bin \citep{Kaastra_16}.

\subsection{Dip count-rate selection}
\label{dip}

The orbital interval $0.95$--$1.12$ samples the dipping phase of 4U~1624--490. Since the main goal of this work is to follow the orbital modulation of the Fe--K absorber velocity, we do not attempt here a detailed reconstruction of the full spectral evolution throughout the dip. The spectra extracted at the lowest count rates are expected to be affected by strong, rapidly variable, and possibly multi-layer obscuration, whose detailed modelling is beyond the scope of this paper. We therefore use the least obscured part of the dip interval to extend the Fe--K absorber velocity curve to the dipping phase, while limiting the impact of the strongest local obscuration.

We considered Observation~2, where the dip phase is well sampled by Resolve, and divided the selected events according to the 2.6--18 keV count rate. The dip-phase analysis discussed below is based on the Resolve data only, so that the count-rate selection and the Fe--K line-centroid measurement are defined consistently from the same high-resolution dataset. Three count-rate selections were initially considered, corresponding to ${\rm CR}>13~{\rm counts~s^{-1}}$, $6<{\rm CR}<13~{\rm counts~s^{-1}}$, and ${\rm CR}<6~{\rm counts~s^{-1}}$, with net exposures of 9.5, 7.0, and 4.2 ks, respectively, as shown in Fig.~\ref{fig:dip_lightcurve}. These selections were used only to identify the least obscured part of the dip.  
In the following spectral analysis, we therefore use only the high-count-rate spectrum, with ${\rm CR}>13~{\rm counts~s^{-1}}$, as an additional Fe--K velocity point close to the dipping phase.

\section{Spectral analysis}
\label{sec:analysis}

\subsection{Baseline continuum modelling: Model A}
\label{sec:model_a}

The spectral analysis was designed to describe, as homogeneously as
possible, the narrow Fe--K absorption features, the underlying
continuum, and the broad Fe--K/broadband curvature throughout the
orbital cycle. For the phase-resolved spectra outside the dip, the
Resolve data were fitted over the 2--18~keV energy range. The only
exception is the 0.32--0.47 phase interval, for which the shorter
effective exposure of the Obs.~2 spectrum causes the source to become
background-dominated above approximately 16~keV. We therefore
restricted the fit to the 2--16~keV band for this phase interval,
because the higher-energy channels do not provide meaningful
constraints on the spectral model. The simultaneous NuSTAR/FPMA and
FPMB spectra were analysed in the 4.5--30~keV band.
All spectral fits were performed with \textsc{xspec} v12.15.0, using
the interstellar abundances of \citet{Wilms00} and the photoelectric
cross-sections of \citet{Verner_96}.

We first adopted a baseline continuum-plus-neutral-absorption model,
hereafter Model~A, of the form
$$
{\tt const*TBabs*TBpcf*(thcomp*bbodyrad)} .
$$
The multiplicative constant accounts for cross-calibration differences
among the instruments. It was fixed at unity for the Resolve spectrum
of Obs.~1 and was allowed to vary for the other spectra included in
each joint fit. For the Resolve-only fit to the high-count-rate dip
spectrum, no cross-calibration constant was required.
The interstellar absorption was described with \texttt{TBabs}, with
the equivalent hydrogen column density fixed at
$N_{\rm H}=4\times10^{22}~{\rm cm^{-2}}$, following
\citet{Xiang_07}. The local neutral absorber was modelled with
\texttt{TBpcf}, whose free parameters are the equivalent hydrogen
column density, $N_{\rm H,pc}$, and the covering fraction, $pcf$.
The intrinsic continuum was described by a Comptonised blackbody
component, \texttt{thcomp*bbodyrad}. The \texttt{thcomp} model
describes thermal Comptonisation \citep{2020MNRAS.492.5234Z} and is
parametrised by the photon index $\Gamma$, the electron temperature of
the Comptonising plasma, $kT_{\rm e}$, and the covering fraction
\texttt{cov\_frac}. The latter was fixed to unity, assuming that the
seed-photon emission is fully Comptonised. The seed spectrum was
described with \texttt{bbodyrad}, which we use as a phenomenological
representation of the soft thermal emission possibly originating from
the neutron-star surface and/or the boundary layer. Its free
parameters are the blackbody temperature, $kT_{\rm bb}$, and the
normalisation, which was used to infer an apparent blackbody radius
assuming a source distance of 15~kpc.

For the phase-resolved fits outside the dip, the photon index $\Gamma$
was allowed to vary independently between Obs.~1 and Obs.~2 within
each orbital-phase interval, while the remaining spectral parameters
were tied between the two observations. The only exception is the
0.87--0.95 interval, in which the values of $\Gamma$ obtained for the
two observations were consistent within their uncertainties and were
therefore tied. The same general continuum framework was applied to
the high-count-rate spectrum extracted from the least obscured part
of the dip, using the Resolve data alone.

Application of Model~A gives statistically unacceptable fits in all
phase intervals. The corresponding fit statistics are
$\chi^2({\rm d.o.f.})=7424.36(4969)$, $3608.84(3396)$,
$3303.30(2993)$, $6347.95(5284)$, $6732.96(4761)$,
$5279.03(3657)$, and $4596.53(2238)$ for the phase intervals from
0.12--0.25 to 0.95--1.12, respectively.
Inspection of the residuals shows strong and structured narrow
absorption features in the Fe--K band, particularly in the
6.6--7.1~keV range. These residuals are associated mainly with
Fe~{\sc xxv} and Fe~{\sc xxvi} absorption, together with additional
higher-energy transitions. Their presence prevents a reliable
determination of the local continuum and of any broader Fe--K
curvature unless the narrow absorption structure is modelled
explicitly. This motivates the introduction of a photoionised
absorber in Model~B. 

\subsection{Photoionised absorption modelling: Model B}
\label{sec:model_b}
\begin{figure*}
\centering
\includegraphics[width=\textwidth]{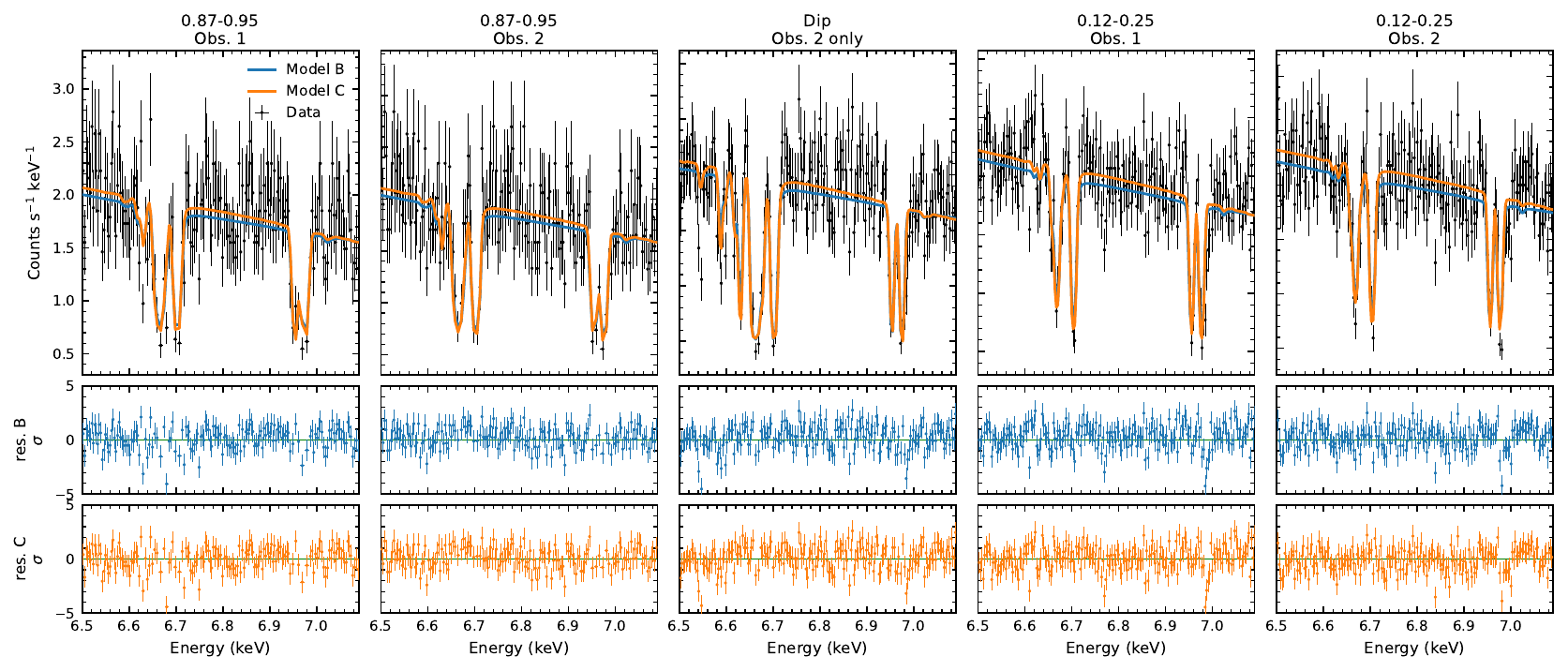}
\caption{Fe--K band spectra for selected orbital-phase intervals. 
For each interval, the upper panel shows the data together with the
best-fitting Model~B and Model~C, while the middle and lower panels show
the residuals in units of $\sigma$ for Model~B and Model~C, respectively.
Model~B already provides a good description of the narrow
Fe~{\sc xxv} and Fe~{\sc xxvi} absorption lines. Structured residuals on
the blue side of the Fe~{\sc xxvi} line are visible in the 0.12--0.25
phase interval, similar to those discussed by \citet{DiazTrigo2026},
although their shape changes between Obs.~1 and Obs.~2.}
\label{fig:feK_modelB_modelC_main}
\end{figure*}

To describe the narrow Fe--K absorption structure, we included the
multiplicative photoionised-absorber component \texttt{warmabs} and
accounted for partial covering of this component using
\texttt{partcov}. The resulting model, hereafter Model~B, is
\begin{equation*}
\begin{aligned}
 & \texttt{const*TBabs*TBpcf}*\texttt{(partcov*warmabs)} \\
&*\texttt{(thcomp*bbodyrad)} \, .
\end{aligned}
\label{eq:modelB}
\end{equation*}
The \texttt{warmabs} component is based on \textsc{xstar}
photoionisation calculations \citep{kallman04} and computes the
transmitted spectrum from pre-calculated population files. Its main
free parameters are the equivalent hydrogen column density of the
ionised gas, $N_{\rm H,WA}$, the ionisation parameter
$\log\xi_{\rm WA}$, the turbulent velocity $v_{\rm turb}$, and a
global redshift parameter $z$, which accounts for the systematic
energy displacement of the absorption features. The ionisation
parameter is defined as $\xi=L_{\rm ion}/(nr^2)$, 
where $L_{\rm ion}$ is the ionising luminosity, $n$ is the gas density,
and $r$ is the distance from the ionising source.

The \texttt{warmabs} component was computed assuming the default
\textsc{xstar} ionising spectral energy distribution, namely an
illuminating power-law continuum with photon index $\Gamma=2$.
Consequently, the inferred ionisation balance is tied to this assumed
SED rather than being calculated self-consistently from the
best-fitting broadband continuum of each orbital-phase interval. The
adopted grid nevertheless provides an adequate phenomenological
description of the narrow Fe--K absorption profiles throughout the
orbital cycle.
The absolute values of $N_{\rm H,WA}$ and $\log\xi_{\rm WA}$ should
therefore be regarded as model-dependent, because a different
illuminating SED could reproduce similar line strengths with a
different combination of column density and ionisation parameter.
Moreover, since the observed continuum changes with orbital phase,
the fitted phase-to-phase variations in $\log\xi_{\rm WA}$ cannot be
interpreted as exact quantitative changes in the physical ionisation
state of the absorber. They should instead be regarded as indicative
variations within the adopted fixed \textsc{xstar} grid, which may
partly reflect the phase-dependent mismatch between the assumed and
actual illuminating continua.

For the same reason, quantitative comparisons of $\log\xi$ with
values reported in other works require caution unless comparable
photoionisation grids, illuminating continua, and broadband spectral
decompositions are adopted. This limitation is not expected to
significantly affect the inferred velocity shifts, which are primarily
constrained by the observed centroids of the absorption lines.

Partial covering of the photoionised absorber was taken into account
by applying \texttt{partcov} to \texttt{warmabs}. This
parametrisation allows only a fraction $f$ of the continuum to be
transmitted through the ionised gas, while the remaining fraction is
left unaffected. Model~B therefore provides an absorption baseline
closely related to the partial-covering framework adopted by
\citet{DiazTrigo2026}. In this phenomenological description, the
uncovered fraction should not be interpreted uniquely: it may
represent geometrical partial covering of an extended continuum
source, but it may also mimic continuum photons scattered into the
line of sight by the ionised plasma.

The inclusion of the partially covering photoionised absorber produces
a clear improvement with respect to Model~A in all phase intervals.
The improvements are $\Delta\chi^2=2120$, 200, 322, 876, 1937, 1387,
and 2135 for the 0.12--0.25, 0.25--0.32, 0.32--0.47, 0.47--0.71,
0.71--0.87, 0.87--0.95, and 0.95--1.12 intervals, respectively. The
corresponding reduced $\chi^2$ values range from approximately 1.0 to
1.1, indicating that Model~B provides an adequate absorption-only
description of the narrow Fe--K features. The best-fitting parameters
obtained with Model~B are reported in
Table~\ref{tab:modelB_all}.

In the 0.25--0.32 and 0.32--0.47 intervals, the turbulent velocity was
fixed at 50~km~s$^{-1}$ because it was not constrained when left free
and systematically converged towards values close to zero. In the
remaining intervals, $v_{\rm turb}$ is modest, with values of
approximately 90--220~km~s$^{-1}$.
The redshift parameter is negative in all phase intervals, showing
that the absorber is systematically blueshifted. The fitted values
span from
$z=(-2.5253^{+0.0002}_{-0.0011})\times10^{-4}$ to
$z=(-14.7\pm0.2)\times10^{-4}$, corresponding to projected
line-of-sight velocities of approximately 75--440~km~s$^{-1}$. The
largest blueshifts are measured in the 0.32--0.47 and 0.47--0.71
intervals, whereas smaller projected velocities are found towards
the dip.

The continuum parameters obtained with Model~B are comparatively
stable. The electron temperature remains within
$kT_{\rm e}\simeq3.17$--3.71~keV, the blackbody temperature varies
between $kT_{\rm bb}\simeq1.02$ and 1.14~keV, and the photon index
spans $\Gamma\simeq2.55$--3.20. The \texttt{bbodyrad} normalisation
corresponds to an apparent blackbody radius of approximately
15--19~km for an assumed distance of 15~kpc.

Figure~\ref{fig:feK_modelB_modelC_main} shows the Fe--K spectra and
residuals for selected orbital-phase intervals. Model~B provides a
good description of the narrow Fe~{\sc xxv} and Fe~{\sc xxvi}
absorption features, which is particularly important because the
absorber velocity is primarily determined from their centroids.
Some structured residuals nevertheless remain on the blue side of
the Fe~{\sc xxvi} feature in the 0.12--0.25 interval, qualitatively
similar to those discussed by \citet{DiazTrigo2026}. In the present
data, however, their shape and relative strength differ between
Obs.~1 and Obs.~2. This indicates that the additional blue-side
structure is not a stable feature and may instead reflect
time-dependent absorber complexity. We therefore do not introduce an
additional absorber component in the phase-resolved fits.

\begin{table*}[ht!]
\centering
\caption{Best--fit parameters  obtained with Model C.}
\label{tab:modelC_all}
\begin{threeparttable}
\setlength{\tabcolsep}{3.0pt}
\renewcommand{\arraystretch}{1.05}
\scriptsize
\begin{tabular}{@{}llccccccc@{}}
\toprule
Component & Parameter
& 0.95--1.12
& 0.12--0.25
& 0.25--0.32
& 0.32--0.47
& 0.47--0.71
& 0.71--0.87
& 0.87--0.95 \\
\midrule

\texttt{const} & Resolve
& --
& \makecell{[1]\\$0.928\pm0.006$}
& \makecell{[1]\\$0.963\pm0.009$}
& \makecell{[1]\\$0.97\pm0.02$}
& \makecell{[1]\\$0.975\pm0.005$}
& \makecell{[1]\\$1.008\pm0.006$}
& \makecell{[1]\\$1.030\pm0.008$} \\

\texttt{const} & FPMA
& --
& \makecell{$1.107\pm0.007$\\--}
& \makecell{$1.128\pm0.009$\\--}
& \makecell{$1.086^{+0.007}_{-0.013}$\\--}
& \makecell{$1.059\pm0.010$\\$1.028\pm0.004$}
& \makecell{--\\$1.073\pm0.006$}
& \makecell{--\\$1.089^{+0.016}_{-0.010}$} \\

\texttt{const} & FPMB
& --
& \makecell{$1.096\pm0.007$\\--}
& \makecell{$1.122\pm0.009$\\--}
& \makecell{$1.078^{+0.005}_{-0.010}$\\--}
& \makecell{$1.044\pm0.010$\\$1.015\pm0.005$}
& \makecell{--\\$1.056\pm0.006$}
& \makecell{--\\$1.068\pm0.010$} \\

\addlinespace[2pt]

\texttt{TBabs} & $N_{\rm H}$
& [4.0]
& [4.0]
& [4.0]
& [4.0]
& [4.0]
& [4.0]
& [4.0] \\

\texttt{TBpcf} & $N_{\rm H,pc}$
& $13.4^{+0.5}_{-0.7}$
& $9.5\pm0.2$
& $7.6^{+1.4}_{-0.3}$
& $10.7\pm0.7$
& $10.3^{+0.2}_{-0.6}$
& $9.2^{+0.2}_{-0.4}$
& $9.4^{+1.1}_{-0.6}$ \\

& $pcf$
& $0.719\pm0.012$
& $0.751^{+0.011}_{-0.019}$
& $0.84^{+0.04}_{-0.06}$
& $0.75\pm0.03$
& $0.711^{+0.014}_{-0.007}$
& $0.740^{+0.022}_{-0.012}$
& $0.776^{+0.057}_{-0.015}$ \\\\

\texttt{PartCov} & $f$
& $0.74\pm0.02$
& $0.69\pm0.02$
& $0.83^{+0.12}_{-0.17}$
& $0.81\pm0.08$
& $0.61^{+0.04}_{-0.02}$
& $0.70\pm0.03$
& $0.66^{+0.03}_{-0.02}$ \\

\addlinespace[2pt]

\texttt{warmabs} & $N_{\rm H,WA}$
& $24\pm3$
& $22\pm2$
& $18^{+19}_{-4}$
& $15^{+3}_{-4}$
& $25^{+4}_{-2}$
& $23^{+3}_{-2}$
& $33^{+5}_{-3}$ \\

& $\log\xi_{\rm WA}$
& $3.14\pm0.03$
& $3.403^{+0.015}_{-0.011}$
& $3.79^{+0.15}_{-0.04}$
& $3.78^{+0.08}_{-0.10}$
& $3.701^{+0.037}_{-0.010}$
& $3.57^{+0.02}_{-0.03}$
& $3.315^{+0.017}_{-0.011}$ \\

& $v_{\rm turb}$ (km\,s$^{-1}$)
& $172\pm18$
& $138\pm12$
& [50]
& [50]
& $115^{+20}_{-17}$
& $225^{+18}_{-13}$
& $227^{+14}_{-26}$ \\

& $z$ ($10^{-4}$)
& $-2.832^{+0.012}_{-0.009}$
& $-5.93^{+0.05}_{-0.02}$
& $-11.2^{+0.3}_{-0.2}$
& $-14.6\pm0.2$
& $-14.34\pm0.05$
& $-9.5\pm0.2$
& $-3.53^{+0.08}_{-0.04}$ \\\\

\texttt{relxillNS} & norm ($10^{-4}$)
& $10.1\pm0.6$
& $7.4\pm0.2$
& $6.5^{+0.5}_{-1.4}$
& $3.9^{+1.0}_{-0.6}$
& $5.9\pm0.2$
& $5.7\pm0.4$
& $6.6\pm0.4$ \\

& $R_{\rm in}$ ($R_{\rm g}$)
& [24.2]
& [53]
& [57]
& [57]
& [105]
& [96]
& [59] \\

& $\log\xi_{\rm refl}$
& $2.50^{+0.06}_{-0.10}$
& $2.63\pm0.04$
& $2.75\pm0.09$
& [2.75]
& $2.74\pm0.04$
& $2.67\pm0.07$
& $2.53^{+0.06}_{-0.09}$ \\

& $kT_{\rm s}$ (keV)
& [=$kT_{\rm e}$]
& [=$kT_{\rm e}$]
& [=$kT_{\rm e}$]
& [=$kT_{\rm e}$]
& [=$kT_{\rm e}$]
& [=$kT_{\rm e}$]
& [=$kT_{\rm e}$] \\

\addlinespace[2pt]

\texttt{thcomp} & $\Gamma$
& $3.47^{+0.14}_{-0.06}$
& \makecell{$3.13\pm0.02$\\$2.92^{+0.02}_{-0.07}$}
& \makecell{$3.01^{+0.22}_{-0.13}$\\$2.65^{+0.11}_{-0.05}$}
& \makecell{$2.58^{+0.02}_{-0.07}$\\$2.90^{+0.07}_{-0.10}$}
& \makecell{$3.363^{+0.008}_{-0.017}$\\$2.967^{+0.012}_{-0.010}$}
& \makecell{$3.27^{+0.05}_{-0.02}$\\$2.949\pm0.015$}
& $3.12^{+0.17}_{-0.05}$ \\

& $kT_{\rm e}$ (keV)
& [2.66]
& $2.649^{+0.012}_{-0.022}$
& $2.73^{+0.07}_{-0.05}$
& $2.96\pm0.05$
& $2.9262^{+0.0007}_{-0.0026}$
& $2.86^{+0.05}_{-0.02}$
& $2.66\pm0.03$ \\

\addlinespace[2pt]

\texttt{bbodyrad} & $kT_{\rm bb}$ (keV)
& [1.05]
& \makecell{$1.04\pm0.02$\\$1.067^{+0.026}_{-0.014}$}
& $1.09\pm0.05$
& $1.012^{+0.013}_{-0.035}$
& $1.096^{+0.005}_{-0.004}$
& $1.073^{+0.026}_{-0.007}$
& $1.05\pm0.06$ \\

& Norm$_{\rm bb}$
& $115.0^{+0.9}_{-1.1}$
& $117.3^{+10.6}_{-1.3}$
& $101^{+14}_{-22}$
& $150^{+22}_{-8}$
& $108.0^{+0.4}_{-2.6}$
& $109^{+2}_{-8}$
& $108^{+6}_{-13}$ \\\\

\texttt{fit} & $\chi^2$ (d.o.f.)
& $2429.01\,(2232)$
& $5171.77\,(4961)$
& $3364.62\,(3390)$
& $2948.46\,(2987)$
& $5282.81\,(5277)$
& $4692.33\,(4754)$
& $3830.19\,(3650)$ \\

& $\Delta\chi^2$
& 33
& 133
& 44
& 32
& 189
& 103
& 62 \\

& F-test c. prob.
& $4.6\times10^{-8}$
& $1.1\times10^{-26}$
& $2.5\times10^{-10}$
& $8.3\times10^{-8}$
& $4.9\times10^{-41}$
& $2.3\times10^{-23}$
& $1.9\times10^{-13}$ \\

& $\sigma$
& 5.3
& 10.6
& 6.3
& 5.4
& 13.4
& 9.9
& 7.3 \\

\bottomrule
\end{tabular}

\begin{tablenotes}[flushleft]
\footnotesize
\item {Notes.}
The dip column refers to the high-count-rate spectrum extracted
from the 0.95--1.12 orbital-phase interval. 
$N_{\rm H}$, $N_{\rm H,pc}$, and $N_{\rm H,WA}$ are given in units of
$10^{22}\ \mathrm{cm^{-2}}$. Parameters reported in square brackets
were held fixed during the spectral fitting. Quoted uncertainties
correspond to the 90\% confidence level. For rows with two entries
in a given orbital-phase interval, the upper and lower
values refer to Obs.~1 and Obs.~2, respectively. $\Delta\chi^2$
values are computed relative to \texttt{Model B}.
\end{tablenotes}
\end{threeparttable}
\end{table*}

More importantly, although Model~B accounts for the dominant narrow
absorption structure, broad and coherent residuals remain in the
Fe--K region and in the associated broadband curvature in several
phase intervals. The broad residuals are most clearly
seen in the joint broadband fit, with NuSTAR providing particularly strong
leverage because of its larger effective area in the Fe--K band and its
broader high-energy coverage. 
Figure~\ref{fig:residuals_modelB_modelC} shows representative residuals obtained   for the 0.12--0.25 and 0.47--0.71 phase intervals. For display purposes only, the residuals were graphically rebinned using the \textsc{xspec} command \texttt{setplot rebin 30 30}. This pattern is difficult to reconcile with a purely narrow absorption-line mismatch. Instead, it suggests that an additional broad emission component is superimposed on the absorption structure. The fact that these deviations are concentrated mainly around the Fe--K band, rather than being randomly distributed over the full fitted energy range, supports the interpretation that \texttt{Model B} is missing a reflection-related component rather than suffering only from an inadequate continuum description.

\subsection{Disc-reflection modelling: Model C}
\label{sec:model_c}

We extended Model~B by adding the ionised relativistic-reflection
component \texttt{relxillNS}. The resulting model, hereafter Model~C,
is
\begin{equation*}
\begin{aligned}
 & \texttt{const*TBabs*TBpcf}*\texttt{(partcov*warmabs)} \\
&*\texttt{[relxillNS+(thcomp*bbodyrad)]} \, .
\end{aligned}
\label{eq:modelC}
\end{equation*}

The \texttt{relxillNS} model belongs to the \texttt{relxill} family
\citep{Garcia2014,Dauser2014}, which combines the angle-dependent
reflection calculations of \texttt{xillver} with a relativistic
convolution kernel accounting for Doppler shifts, gravitational
redshift, and light bending. In contrast to reflection models
illuminated by a power-law continuum, \texttt{relxillNS} adopts a
thermal illuminating spectrum appropriate for neutron-star systems.
This makes it suitable for low-mass X-ray binaries in which the
primary emission is produced by a relatively low-temperature,
optically thick Comptonising region or boundary layer.

In this framework, the reflected spectrum includes both the Fe--K
line complex and the associated broadband continuum. Model~C
therefore provides a physically motivated description of the broad
Fe--K curvature rather than representing it with an isolated
phenomenological emission line.
The presence of a broad Fe--K emission component is also consistent with previous observations of 4U~1624$-$490. The XMM-Newton spectrum analysed by \citet{Parmar_02} showed an Fe emission line centred at $\sim6.58$~keV with $\sigma \simeq 470$~eV, while the Chandra/HETGS observation studied by \citet{Iaria2007b} revealed a broad Fe emission line at $\sim6.64$~keV with FWHM $\sim700$~eV. More recently, \citet{Gnarini_24} found that the inclusion of a reflection component is required to remove broadened Fe-line residuals between 6 and 7~keV in simultaneous NICER and NuSTAR spectra. Therefore, {Model~C} represents a physically motivated extension of {Model~B}, aimed at testing whether the broadband data require reflection in addition to the phase-dependent ionised absorption.

In the reflection fits, the inclination was fixed at $64^\circ$,
consistent with the high-inclination geometry inferred for this
dipping source, and the emissivity profile was assumed to scale as
$r^{-3}$. The outer disc radius was fixed at
$R_{\rm out}=900~r_{\rm g}$, and the dimensionless spin parameter was
fixed at $a=0$, because the reflection spectrum at the inferred inner
radii is not expected to depend significantly on the exact value of
the compact-object spin. For each phase-resolved spectrum, $R_{\rm in}$ was fixed to the
best-fitting value obtained from preliminary fits in order to reduce
degeneracies in the final modelling. We define the reflected fraction
as the ratio between the 0.1--100~keV flux associated with
\texttt{relxillNS} and that of the incident continuum described by
\texttt{thcomp*bbodyrad}.

The disc density was fixed at $\log N=18$, corresponding to
$N=10^{18}~{\rm cm^{-3}}$, as a fiducial value representative of a
dense neutron-star accretion disc and to reduce the number of weakly
constrained parameters. The temperature of the illuminating photons
was tied to the electron temperature of the Comptonising continuum.
This choice is motivated by the fact that the primary Comptonised
spectrum is close to the saturated regime and is therefore
approximately Wien-like, with a characteristic temperature of the
order of $kT_{\rm e}$.

\begin{figure}[h]
\centering
\includegraphics[width=0.95\columnwidth]{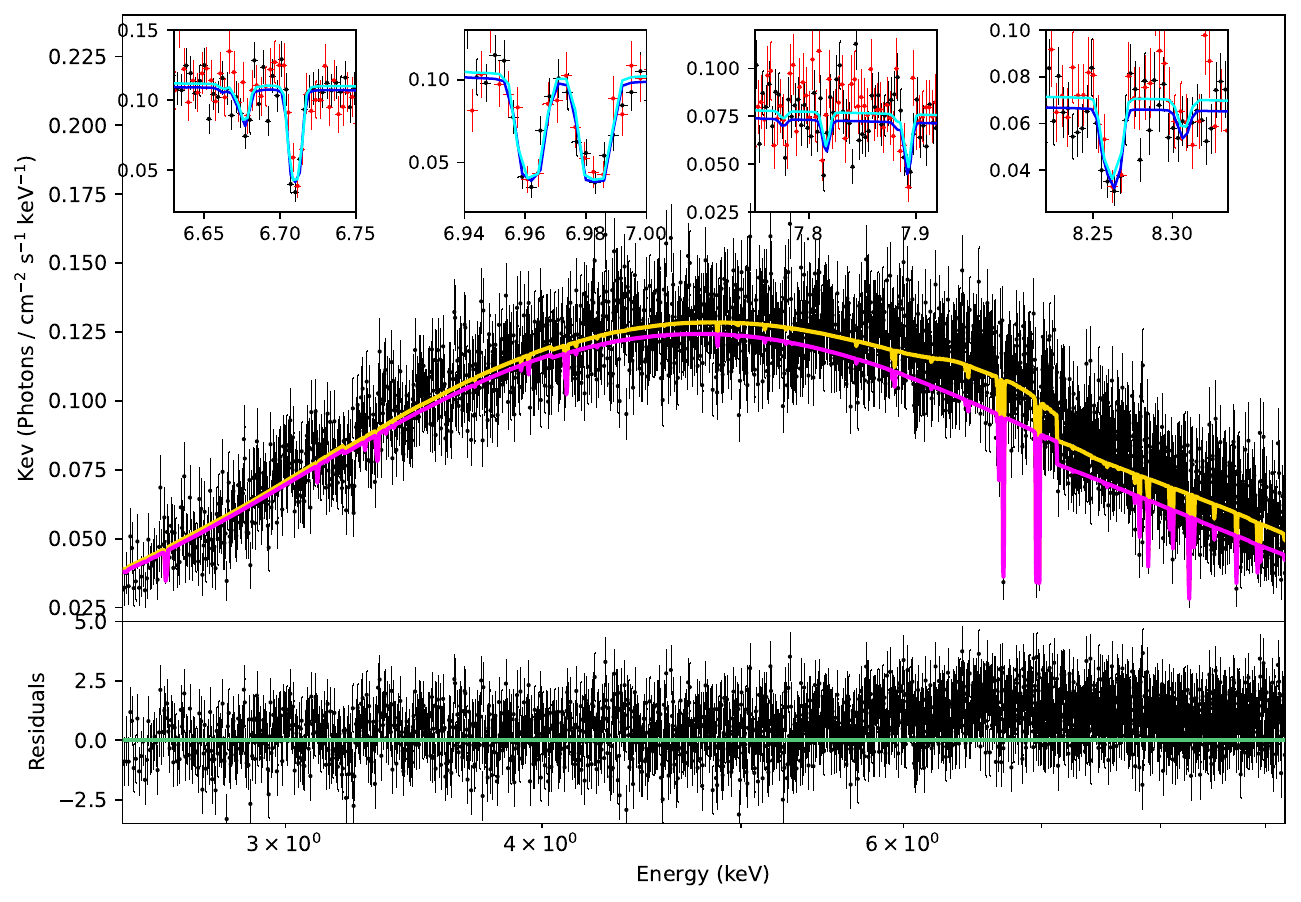}
\caption{ Representative XRISM/Resolve spectrum for the orbital-phase interval 0.47--0.71. For clarity, the main panel shows only Obs.~1, while the insets show zooms on the main absorption complexes for both Obs.~1 and Obs.~2. The lower panel shows the residuals obtained after excluding the disc-reflection component from the best-fitting {Model~C}.}
\label{fig:modelC_reflection_diagnostic}
\end{figure}

The inclusion of Model~C produces a statistically significant
improvement in all seven phase-resolved datasets.
We also apply the reflection model to intervals where
Model~B already yields a reduced $\chi^2$ close to,
or marginally below, unity, to maintain a consistent
spectral framework across orbital phase.
The best-fitting
parameters are reported in Table~\ref{tab:modelC_all}. The associated
$\Delta\chi^2$ values are large despite the addition of only a small
number of free parameters, and the corresponding chance probabilities
are negligible. This indicates that the broad Fe--K and broadband
residuals are effectively accounted for by the reflection component.

The continuum parameters remain relatively stable throughout the
orbit. The blackbody temperature remains close to
$kT_{\rm bb}\simeq1$~keV, while the electron temperature of the
Comptonising plasma lies in the range
$kT_{\rm e}\simeq2.6$--3.0~keV. The local neutral absorber shows
moderate variability outside the dip, with
$N_{\rm H,pc}\simeq(0.8$--$1.1)\times10^{23}~{\rm cm^{-2}}$ and
covering fractions of approximately 0.7--0.8. In the high-count-rate
dip spectrum, the neutral column increases to
$N_{\rm H,pc}=1.34^{+0.05}_{-0.07}\times10^{23}~{\rm cm^{-2}}$,
while its covering fraction remains comparable to the out-of-dip
values.

The fitted warm-absorber column density is of the order of a few
$10^{23}~{\rm cm^{-2}}$. 
Within the adopted fixed \textsc{xstar} grid, the fitted
$\log\xi_{\rm WA}$ values are higher at intermediate phases and
lower close to the dip. As discussed in Sect.~\ref{sec:model_b}, however, these
variations are model-dependent and should not be interpreted as
exact quantitative changes in the physical ionisation state.

The redshift parameter displays a coherent phase dependence. The
absorber is blueshifted at all orbital phases, with the largest
projected velocities measured at intermediate phases and progressively
smaller blueshifts towards the dip. For the high-count-rate dip
spectrum, we obtain
$z=(-2.832^{+0.012}_{-0.009})\times10^{-4}$, corresponding to
$v_{\rm los}\simeq-85~{\rm km~s^{-1}}$; this value is adopted for the dip-phase point in the orbital-modulation
analysis.

\begin{figure*}[ht!]
    \centering    \includegraphics[width=0.95\textwidth]{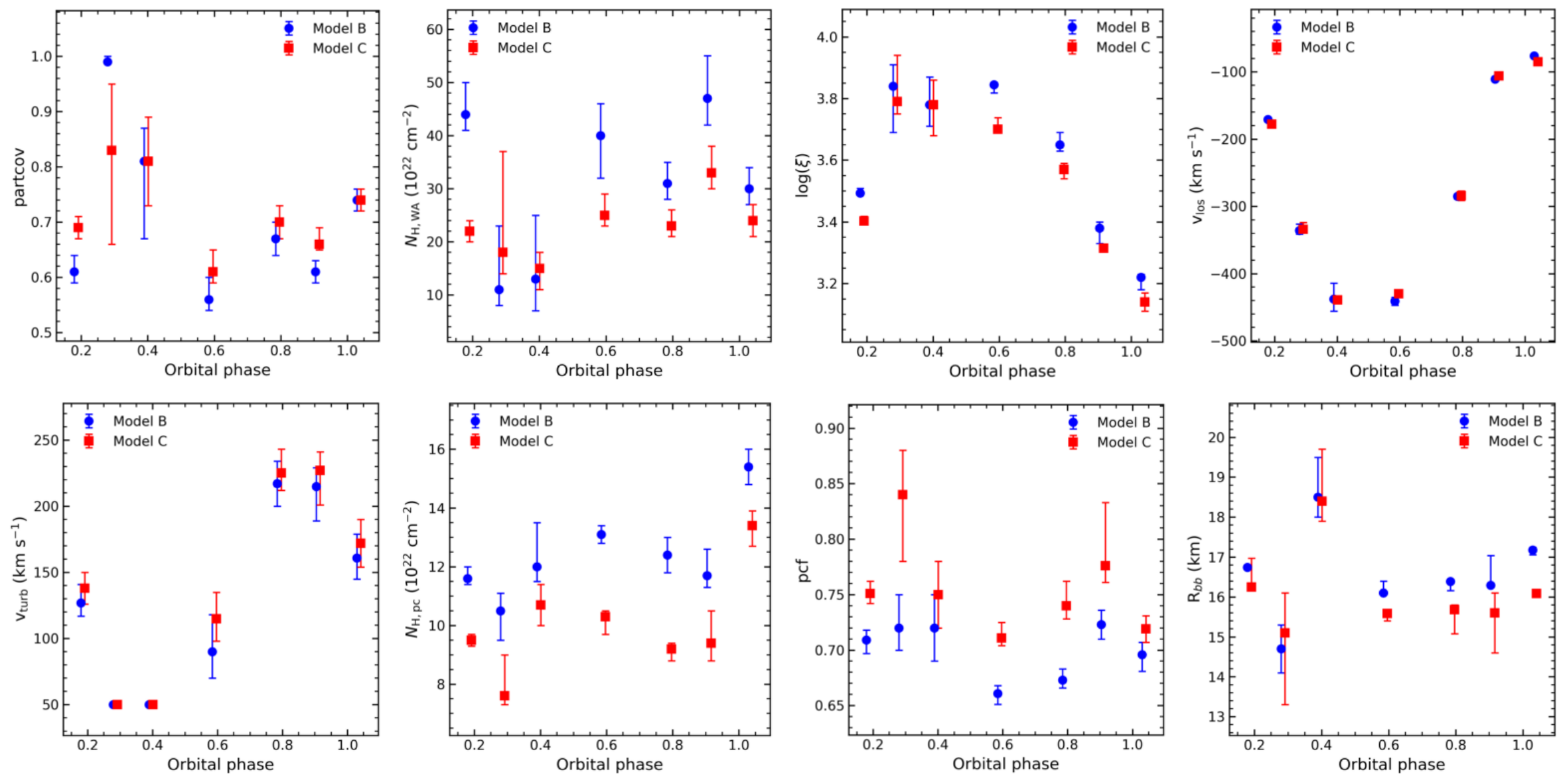}
    \caption{
Comparison between the absorber and partial-covering parameters obtained with Model~B and Model~C as a function of orbital phase. Blue circles and red squares indicate Model~B and Model~C, respectively. The panels show the warm-absorber covering fraction, $N_{\rm H,WA}$, $\log\xi$, $v_{\rm los}$, $v_{\rm turb}$, $N_{\rm H,pc}$, $pcf$ and R$_{\rm bb}$. Error bars correspond to the 90\% confidence level.    }
\label{fig:modelB_modelC_parameters}
\end{figure*}

Figure~\ref{fig:modelC_reflection_diagnostic} illustrates the spectral
decomposition obtained with Model~C for the representative
0.47--0.71 phase interval. The reflection component primarily accounts
for the broad curvature across the Fe--K band and the surrounding
continuum. The narrow absorption troughs are reproduced at the
observed energies in both Obs.~1 and Obs.~2, showing that reflection
improves the description of the underlying broad emission without
artificially shifting the centroids of the Fe~{\sc xxv} and
Fe~{\sc xxvi} absorption lines. The narrow absorption troughs are reproduced at the same energies
in Models B and C. As shown in Figure~\ref{fig:modelB_modelC_parameters}  and quantified in Tab.\ref{tab:rbb_v_phase},
the inclusion of reflection does not significantly alter the
line-of-sight velocities of the photoionised absorber.

Some model dependence is instead observed in the inferred absorber
column densities and partial-covering parameters. In particular,
$N_{\rm H,WA}$ is generally lower in Model~C than in Model~B. Once
reflection is included, part of the broadband curvature previously
absorbed into the photoionised-absorber description is instead
accounted for by the reflected component. A similar behaviour is
observed for $N_{\rm H,pc}$, which also tends to decrease when
reflection is included.

The fitted ionisation parameter and absorber column density show some
model dependence, owing to the different decomposition of the broadband
spectral curvature. By contrast, the turbulent velocities obtained with
Models~B and C are mutually consistent within their uncertainties and
primarily show a phase-dependent increase near and during the dip.
 Since Model~C provides the statistically and physically preferred
description of the broadband spectrum, we adopt it for the subsequent
radial-velocity analysis.

\begin{table}
\caption{Comparison of the line-of-sight velocities obtained with Model~B and Model~C.}
\label{tab:rbb_v_phase}
\centering
\tiny
\setlength{\tabcolsep}{4pt}

\begin{tabular}{lcccc}
\hline\hline
 & Model B & \multicolumn{3}{c}{Model C}\\
Phase range
& $v_{\rm los}$ ($\mathrm{km\,s^{-1}}$)
& $R_{\rm bb}$ (km)
& $v_{\rm los}$ ($\mathrm{km\,s^{-1}}$)
& $\Delta v_{\rm C-B}$ ($\mathrm{km\,s^{-1}}$)\\
\hline
0.12--0.25
& $-170.9^{+1.2}_{-0.6}$
& $16.25_{-0.09}^{+0.72}$
& $-177.7^{+1.9}_{-0.8}$
& $-6.8^{+2.0}_{-1.4}$ \\

0.25--0.32
& $-336^{+10}_{-5}$
& $15.1_{-1.8}^{+1.0}$
& $-334^{+10}_{-5}$
& $+2.0^{+11.2}_{-11.2}$ \\

0.32--0.47
& $-438^{+24}_{-18}$
& $18.4_{-0.5}^{+1.3}$
& $-439^{+6}_{-5}$
& $-1.0^{+19.0}_{-24.5}$ \\

0.47--0.71
& $-441\pm6$
& $15.59_{-0.19}^{+0.03}$
& $-429.8^{+1.1}_{-1.4}$
& $+11.2^{+6.1}_{-6.2}$ \\

0.71--0.87
& $-285^{+2}_{-4}$
& $15.68_{-0.60}^{+0.12}$
& $-284\pm7$
& $+1.0^{+8.1}_{-7.3}$ \\

0.87--0.95
& $-110.7^{+0.2}_{-0.5}$
& $15.6_{-1.0}^{+0.5}$
& $-105.8^{+2.4}_{-1.3}$
& $+4.9^{+2.5}_{-1.3}$ \\

0.95--1.12
& $-76.4\pm0.9$
& $16.09_{-0.08}^{+0.05}$
& $-84.9\pm0.3$
& $-8.5\pm1.0$ \\
\hline\hline
\end{tabular}

\tablefoot{
The last column gives
$\Delta v_{\rm C-B}
= v_{\rm los}(\mathrm{Model~C})
- v_{\rm los}(\mathrm{Model~B})$.
$R_{\rm bb}$ is computed from the \texttt{bbodyrad} normalisation
assuming $D=15$~kpc.
$v_{\rm los}$ is derived from the measured $z$ parameter assuming
the non-relativistic approximation $v_{\rm los}\simeq zc$.
The blackbody radius reported in the third column is that obtained
with Model~C.
The tabulated velocity uncertainties are statistical only.
The Resolve absolute energy-scale uncertainty, equivalent to
approximately $12.9~\mathrm{km\,s^{-1}}$ at Fe\,{\sc xxvi}
Ly$\alpha$, is not included in the individual tabulated values but
is added in quadrature when fitting the orbital velocity modulation
in Sect.~\ref{sect:radial_velocity}.
}

\end{table}

\section{Discussion}
\label{sec:discussion}

We adopt Model~C as the baseline spectral description and examine the
orbital modulation of the absorber velocity, its possible dynamical
implications, and the evolutionary nature of the system.

\subsection{Orbital modulation of the absorber line-of-sight velocity}
\label{sect:radial_velocity}

\begin{figure}
    \centering
    \includegraphics[width=\columnwidth]{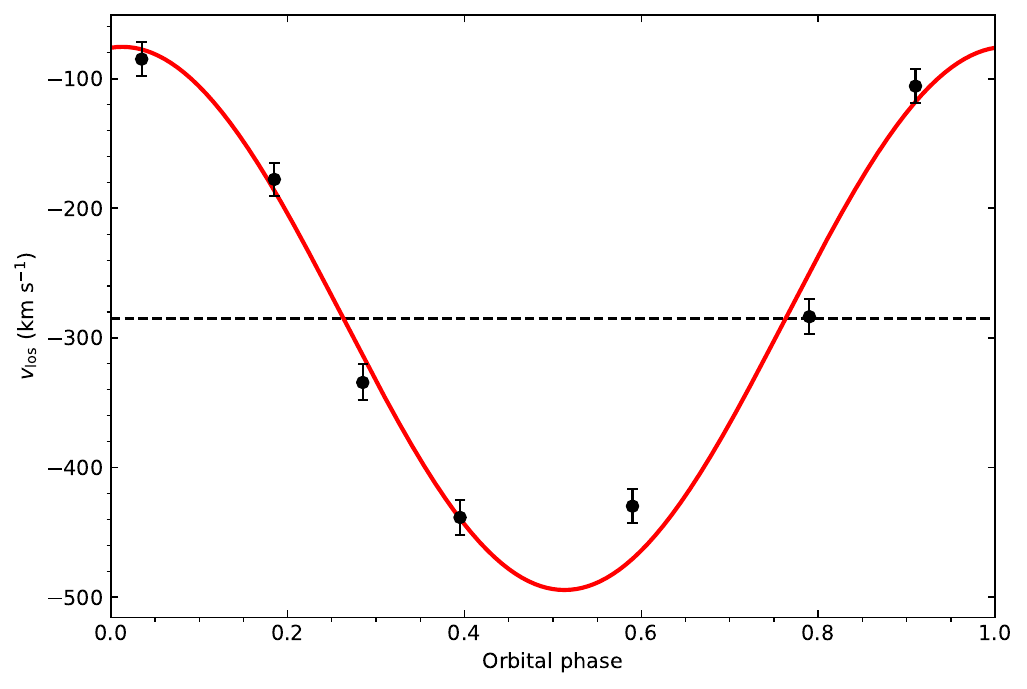}
    \caption{Line-of-sight velocity of the ionised absorber as a function of orbital phase together with the best-fitting sinusoidal model (red curve). The  dashed line marks the constant term $v_{0}$. Phase zero is defined at the dip centre.}
    \label{fig:vlos_fit}
\end{figure}

The line-of-sight velocity of the ionised absorber was estimated from
the Doppler shift of the absorption lines, parameterised in the
spectral fits through the redshift parameter $z$ of
\texttt{warmabs}. Since the measured values of $|z|$ are small, we
adopted the non-relativistic relation
$ v_{\rm los}=zc$,
where $c$ is the speed of light. The resulting velocities are reported
in Table~\ref{tab:rbb_v_phase} and shown in
Fig.~\ref{fig:vlos_fit}.
For the 0.95--1.12 dip interval, we used the velocity measured from the
high-count-rate spectrum defined in Sect. \ref{dip}.
 It provides
the most suitable dip-phase estimate of the centroid displacement of
the narrow Fe--K absorption features.

We modelled the phase dependence of the absorber velocity using the
sinusoidal function $    v(\phi)=v_0+
    K\sin\left[2\pi(\phi-\phi_0)\right], $
where $v_0$ is the phase-averaged velocity offset, $K$ is the
semi-amplitude of the modulation, and $\phi_0$ is its phase offset.

Each velocity measurement was obtained from a spectrum accumulated
over a finite orbital-phase interval. The measured velocity should
therefore not be compared simply with the value of the sinusoidal
model evaluated at the centre of the corresponding phase bin.
Instead, for each interval $[\phi_{i,\rm lo},\phi_{i,\rm hi}]$, we compared
the measured velocity $v_i$ with the model prediction averaged over the
full phase bin:
\begin{equation}
v_{{\rm mod},i} =
\frac{1}{\Delta\phi_i}
\int_{\phi_{i,\rm lo}}^{\phi_{i,\rm hi}}
\left\{
v_0 + K\sin\left[2\pi(\phi-\phi_0)\right]
\right\}\,{\rm d}\phi ,
    \label{eq:bin_averaged_velocity}
\end{equation}
where $    \Delta\phi_i=\phi_{i,\rm hi}-\phi_{i,\rm lo} $.
The parameters $v_0$, $K$, and $\phi_0$ were left free and determined
simultaneously by direct minimisation of the $\chi^2$ statistic comparing
$v_i$ with $v_{{\rm mod},i}$.
The finite phase-bin width is thus incorporated directly into the
model prediction and is not treated as an independent uncertainty on
the phase coordinate.

The statistical uncertainties on $v_{\rm los}$, originally evaluated
at the 90\% confidence level, were converted to Gaussian-equivalent
$1\sigma$ uncertainties. The asymmetric lower and upper uncertainties
were retained in the evaluation of the fit statistic.

We also included the systematic uncertainty associated with the
Resolve absolute energy scale. We adopted $ \Delta E_{\rm sys}=0.3\ {\rm eV}$\footnote{See Sect.~5 and Table~5.1 of the
\href{https://heasarc.gsfc.nasa.gov/docs/xrism/proposals/POG/Resolve.html}
{XRISM Proposers' Observatory Guide}.}
At the energy of the Fe\,{\sc xxvi} Ly$\alpha$ transition, this
corresponds to a velocity uncertainty of $12.9\ {\rm km\ s^{-1}} $.
For the fit, this contribution was added in quadrature to the
statistical lower and upper uncertainty of each velocity measurement.
The corresponding fit statistic is
$\chi^2/{\rm d.o.f.}=13.264/4$, indicating residual scatter beyond
that expected from the assigned uncertainties.

To account conservatively for this residual dispersion, we rescaled
the parameter uncertainties by the factor $\sqrt{\chi^2_{\rm red}}$. 
This prescription corresponds to assuming that the unexplained
dispersion can be represented, to first order, by a common
multiplicative underestimation of the velocity uncertainties. The
rescaling does not change the location of the $\chi^2$ minimum and
therefore leaves the best-fitting parameter values unchanged. The
resulting parameters and rescaled $1\sigma$ uncertainties are
$v_0=-285\pm9\ {\rm km\ s^{-1}}$, 
$K=210\pm14\ {\rm km\ s^{-1}}$, and $ \phi_0=0.763\pm 0.010$.

A key question is whether the inferred semi-amplitude is driven by
the phase intervals in which the absorber is expected to be most
complex. This is particularly relevant for the $0.12$--$0.25$
interval, corresponding to the post-dip/p0 interval discussed by
\citet{DiazTrigo2026}, and for the $0.95$--$1.12$ dip interval,
although only its high-count-rate spectrum was used in our analysis.
We therefore repeated the direct minimisation after excluding both
measurements, adopting the same bin-averaged sinusoidal model and the
same treatment of the statistical and energy-scale uncertainties.

The five-point fit gives a $ \chi^2/{\rm d.o.f.}=11.153/2$. After rescaling the profile uncertainties by $\sqrt{\chi^2_{\rm red}}$, the
five-point solution becomes 
$v_0=-290\pm 15\ {\rm km\ s^{-1}}$ , $K=200\pm 26\ {\rm km\ s^{-1}}$, and $\phi_0=0.76\pm0.02$.
 
The five-point solution is consistent with the seven-point fit within
the rescaled uncertainties, demonstrating that the inferred modulation
is not driven solely by the dip and post-dip measurements. The elevated
$\chi^2_{\rm red}$ nevertheless indicates that a sinusoid provides only
a first-order description and does not exclude unresolved absorber
complexity in these intervals.

If, to first order, the observed velocity modulation traces the
orbital motion of the neutron star, superior conjunction occurs at
$\phi\simeq0.76$. This implies an offset of approximately
$\Delta\phi\simeq0.2$ with respect to the dip centroid. Such an offset
is physically plausible in dipping low-mass X-ray binaries, since the
dip is generally associated with absorption in the azimuthally
displaced stream--disc impact region rather than directly with the
donor star itself
\citep{WhiteHolt1982,Frank1987}.

Our results are qualitatively consistent with the recent XRISM
analysis of the same dataset by \citet{DiazTrigo2026}. Both analyses
recover a phase-dependent velocity modulation of the highly ionised
absorber that can be described, to first order, by a sinusoidal trend.
The main difference concerns the treatment of the absorber complexity
close to the dip. Our seven-point fit yields a velocity semi-amplitude
of approximately $210\ {\rm km\ s^{-1}}$ and a phase-averaged velocity
offset of approximately $-285\ {\rm km\ s^{-1}}$.

As an additional stress test, we independently shifted the
velocities of the dip and post-dip measurements by
$\pm100$ and $\pm150\ {\rm km\,s^{-1}}$ and repeated the fit for
all 16 combinations. The resulting semi-amplitudes span
$K\simeq147$--$282\ {\rm km\,s^{-1}}$. Although these deliberately
large perturbations substantially broaden the allowed range of
$K$, none removes the non-zero modulation implied by the remaining
measurements.

This range illustrates the sensitivity of the fitted
semi-amplitude to the imposed velocity perturbations,
rather than providing a confidence interval or a
calibrated estimate of the uncertainty associated with
a second absorber. Although the modulation persists
under these perturbations, its amplitude, and hence
the inferred donor mass under an orbital interpretation,
remain sensitive to unresolved absorber complexity.

\subsection{Constraints on the donor star from the observed velocity amplitude}
\label{sec:donor_implications}

If the observed line-of-sight velocity modulation is interpreted, to first order, as tracing the orbital motion of the neutron star, the measured semi-amplitude implies a donor mass function
\[
f(M_2)=\frac{P_{\rm orb}K_1^3}{2\pi G}
      =\frac{(M_2\sin i)^3}{(M_1+M_2)^2},
\]
where $K_1$ is the radial-velocity semi-amplitude of the neutron star, $M_1$ is the neutron-star mass, $M_2$ is the donor mass, and $i$ is the orbital inclination. Adopting $P_{\rm orb}=0.869896(1)$~d \citep{Liao_15} and $K_1=210\pm14 $~km~s$^{-1}$, we obtain
\[
f(M_2)=0.83\pm 0.17\,M_\odot.
\]
Assuming a canonical neutron-star mass of $M_1=1.4\,M_\odot$ and an inclination of $i=64^\circ$, this corresponds to a donor mass of $M_2 = 2.66\pm0.32\,M_\odot$. The corresponding orbital separation from Kepler's third law is $ a = \left(4.26\pm0.11\right)\times10^{11}\ {\rm cm}$,
while the Eggleton approximation gives Roche-lobe radii
\[
R_{{\rm L},1}=\left(1.383\pm0.006\right)\times10^{11}\ {\rm cm}
\]
and $ R_{{\rm L},2}=\left(1.85\pm 0.10\right)\times10^{11}\ {\rm cm}$.
Taken at face value, this solution points to a relatively massive donor. In terms of standard main-sequence classifications, a star of a few solar masses is naturally compatible with a late B-type dwarf, approximately of spectral type B9\,V--B9.5\,V \citep{Eker2018, Pecaut2013}.

A useful consistency check is provided by the near-infrared properties of the proposed counterpart identified by \citet{Wachter2005}, who reported an apparent magnitude of $m_{K_{\rm s}}=18.3\pm0.1$ mag. Using the interstellar column density inferred from the dust-scattering analysis, $N_{\rm H}=4\times10^{22}\,\mathrm{cm^{-2}}$ \citep{Xiang_07}, together with the empirical $N_{\rm H}$--$A_V$ relation of \citet{Guver2009} and a standard extinction law \citep{Cardelli1989}, we derive $A_V = 18.1\pm0.7$ mag and hence $A_{K_{\rm s}} = 2.12\pm0.08$ mag. For a source distance of $15 \pm 3$ kpc \citep{Xiang_07}, the distance modulus is $\mu = 15.88 \pm 0.43$ mag. Adopting the dwarf calibration of \citet{Pecaut2013}, we then obtain expected apparent magnitudes of  $m_{K_s}(\mathrm{B9\,V}) = 18.6 \pm 0.4$ mag, and $m_{K_s}(\mathrm{B9.5\,V}) = 18.7 \pm 0.4$ mag, all consistent with the observed infrared counterpart. This agreement does not prove the donor classification, since the near-infrared flux may include non-stellar contributions and part of the absorbing column is local to the binary, but it does show that a late B-type donor is not excluded on simple photometric grounds.

An additional consistency check comes from Roche geometry. Modern calibrations indicate radii of roughly $\sim2.5\,R_\odot$ for B9\,V--B9.5\,V dwarfs \citep{Pecaut2013,Eker2018}. Using the Roche equipotential passing through $L_1$ for $M_2=2.66\pm0.32 \,M_\odot$, we estimate a donor polar radius of $r_{\rm pole}=2.48\pm0.15\,R_\odot$ which is  fully consistent with a Roche-lobe-filling late B-type companion. In this sense, the dynamical inference from the absorber-velocity curve, the infrared brightness of the proposed counterpart, and the Roche-lobe geometry all converge on the same donor-mass regime, thereby lending physical consistency to the dynamical interpretation.

If this interpretation is correct, the excess neutral absorption is unlikely to be supplied by a standard radiatively driven stellar wind, which is expected to be weak in late-B main-sequence stars. It is more naturally explained as arising within the accretion environment itself, for example, in the outer disc rim, the stream--disc impact region, or a dense disc atmosphere/outflow.

This solution is also geometrically consistent with the absence of X-ray eclipses. Since 4U~1624--490 is a dipping source, the system must be viewed at high inclination; however, the donor should not occult the compact object at superior conjunction. Approximating the relevant transverse size of the Roche-lobe-filling donor by its polar Roche-lobe radius, the no-eclipse condition can be written as
$$
i < i_{\rm max}=\arccos\left(\frac{r_{\rm pole}}{a}\right),
$$
where $a$ is the binary separation and $r_{\rm pole}$ is the polar radius of the donor Roche lobe, we obtain $i_{\rm max}=66\pm 1$  deg. The inclination adopted in the reflection fits, $i=64^\circ$, is therefore compatible with the nominal no-eclipse limit. Thus, the dynamical solution, the Roche-lobe geometry, the proposed near-infrared counterpart, and the absence of eclipses are mutually consistent within the uncertainties.

 The above interpretation is subject to two distinct
limitations. First, unresolved absorption components
may bias the measured velocity semi-amplitude. For
example, the lower value obtained in our perturbation
test, $K\simeq147$~km~s$^{-1}$, would imply
$f(M_2)\simeq0.29\,M_\odot$ and
$M_2\simeq1.49\,M_\odot$ for
$M_1=1.4\,M_\odot$ and $i=64^\circ$, if identified
with the neutron-star orbital semi-amplitude.
Thus, substantially lower donor masses are possible
even within an orbital interpretation.

 Second, correctly disentangling an absorption component
does not by itself establish that its velocity modulation
traces the neutron-star orbital motion. The selected
plasma may be located elsewhere in the binary, and its
velocity field and the phase-dependent viewing geometry
may contribute to the observed modulation. Unless these
contributions can be separated from the neutron-star
orbital motion, the measured semi-amplitude cannot be
identified directly with $K_1$, and the mass-function
argument does not provide a secure donor-mass constraint.

A low-mass companion therefore remains possible.
As discussed in Appendix~\ref{app:lowmass_donor},
a representative $0.5\,M_\odot$ donor would need
a radius of approximately $1.4\,R_\odot$ to fill
its Roche lobe at the 20.9~hr orbital period,
requiring an evolved, inflated, or stripped star
rather than an unevolved main-sequence companion.
Its interpretation would also require accounting
for the contribution of the accretion flow and/or
X-ray irradiation to the observed $K_{\rm s}$-band
flux. These considerations constrain the nature
of a possible low-mass donor but do not exclude it.

\subsection{Possible intermediate-mass nature and accretion geometry}

If the donor mass inferred above is correct, 4U~1624--490
is more naturally classified as an intermediate-mass X-ray binary
than as a classical low-mass system
\citep{Tauris2006,Podsiadlowski2002}.
 In this interpretation, the donor is a
roughly $2.7\,M_\odot$ Roche-lobe-filling late B-type star.

 A late B-type donor filling its Roche lobe would be expected to respond on a relatively short thermal timescale, naturally leading to rapid and highly non-conservative mass transfer. This provides a plausible framework in which the strong local absorption and complex dipping phenomenology may be linked to dense material in the accretion environment. A quantitative order-of-magnitude discussion of the thermal-timescale evolution is given in Appendix~\ref{app:imxb_evolution}.
If the mass transfer is strongly non-conservative, part of the expelled material could also contribute to the local neutral absorption inferred from the X-ray spectra. A simple order-of-magnitude estimate shows that a circumbinary reservoir could in principle provide the observed column density, and is presented in Appendix~\ref{app:circumbinary_absorber}.

The absorber properties are compatible with a dense Compton-heated
atmosphere or wind base. The measured blueshifts favour outward motion,
but do not by themselves establish that the plasma is part of an
unbound large-scale wind; the relevant physical scales are discussed
in Appendix~\ref{app:compton_wind}.

The reflection component also contributes to the physical picture.
Its orbital visibility appears to vary, suggesting that reflection and
absorption may be coupled manifestations of the same high-inclination
accretion geometry. A more detailed discussion is given in Appendix~\ref{app:reflection}.

\section{Conclusions}

We have presented a phase-resolved XRISM/Resolve and NuSTAR study
of the dipping neutron-star X-ray binary 4U~1624--490. The high
spectral resolution of Resolve in the Fe--K band, combined with the
broadband coverage of NuSTAR, allows us to follow the orbital
evolution of the ionised absorber and to test the role of disc
reflection.

The main results are as follows. First, the Fe~{\sc xxv} and
Fe~{\sc xxvi} absorption features are detected over most orbital
phases and show strong changes in column density, ionisation state,
covering fraction, and velocity shift. This confirms that the line of
sight intercepts a complex, azimuthally structured plasma associated
with the outer disc, stream--disc impact region, and/or disc
atmosphere.

Second, absorption alone does not provide a complete description of
the spectra. A model including neutral and ionised absorption
reproduces the narrow Fe--K lines, but leaves broad residuals in the
Fe--K band. Adding an ionised reflection component significantly
improves the fits and accounts for this broad curvature without
shifting the centroids of the narrow absorption lines.

Third, the ionised absorber is systematically blueshifted and its
line-of-sight velocity shows a coherent orbital modulation. The
best-fitting sinusoid gives a semi-amplitude
$K=210\pm14$~km~s$^{-1}$, corresponding to a mass function
$f(M)=0.83\pm0.17\,M_\odot$ if the modulation traces the
orbital motion of the neutron star. This velocity modulation remains
consistent when the dip and post-dip intervals are excluded, showing
that it is not driven only by the most complex absorber phases.
Residual structure in the blue wing of Fe\,{\sc xxvi} may indicate
additional kinematic or ionisation complexity during and close to
the dip. The present spectra do not uniquely require a second
independently constrained absorber, but unresolved multi-component
absorption remains a possible source of systematic uncertainty in
these phase intervals.

If interpreted dynamically, the measured velocity amplitude implies a
donor mass of roughly $2.66\pm0.32\,M_\odot$ for plausible
high-inclination solutions, compatible with a Roche-lobe-filling late
B-type companion and with the proposed near-infrared counterpart.
This would place 4U~1624--490 more naturally among intermediate-mass
X-ray binaries than among classical low-mass systems. 
However, the range $K\simeq147$--$282$~km~s$^{-1}$
obtained in our perturbation test illustrates the
sensitivity of the inferred donor mass to unresolved
absorber complexity and permits substantially lower
masses than the nominal late-B solution. Moreover,
even a correctly disentangled absorption component
need not trace the neutron-star orbital motion,
because its velocity field and viewing geometry
may contribute to the observed modulation.
A low-mass companion therefore cannot be excluded.

Future high-resolution phase-resolved observations and independent
constraints on the donor star will be needed to establish whether the
absorber velocity curve is a true dynamical tracer and whether the
ionised plasma is best interpreted as a disc atmosphere, a failed
wind, or a genuine outflow.

\begin{acknowledgements}
R.I. and T.D.S. acknowledge support from PRIN-INAF
2019 with the project  ``Probing the geometry of accretion:
from theory to observations'' (PI: Belloni).  
FB acknowledges support from INAF Large Grant 2023 BLOSSOM F.O. 1.05.23.01.13. C.M. acknowledges  support from the Spanish MINECO (PID2023-148661NB-I00). 
This research has made use of data and/or software provided by the High Energy Astrophysics Science Archive Research Center (HEASARC), which is a service of the Astrophysics Science Division at NASA/GSFC.
\end{acknowledgements}

\bibliographystyle{bibtex/aa}
\bibliography{bibtex/bibliography}{}

@article{WhiteHolt1982,
  author       = {White, N. E. and Holt, S. S.},
  title        = {Accretion disk coronae},
  journal      = {ApJ},
  year         = {1982},
  volume       = {257},
  pages        = {318--337},
  doi          = {10.1086/160994}
}

@article{Frank1987,
  author  = {Frank, J. and King, A. R. and Lasota, J.-P.},
  title   = {The light curves of low-mass X-ray binaries},
  journal = {A\&A},
  year    = {1987},
  volume  = {178},
  pages   = {137--142}
}

@article{ChurchBalucinska2004,
  author  = {Church, M. J. and Ba{\l}uci{\'n}ska-Church, M.},
  title   = {Measurements of accretion disc corona size in LMXB},
  journal = {MNRAS},
  year    = {2004},
  volume  = {348},
  pages   = {955--966}
}

@article{Boirin2005,
  author  = {Boirin, L. and M{\'e}ndez, M. and D{\'i}az Trigo, M. and Parmar, A. N. and Kaastra, J. S.},
  title   = {A highly-ionized absorber in the X-ray binary 4U 1323-62: A new explanation for the dipping phenomenon},
  journal = {A\&A},
  year    = {2005},
  volume  = {436},
  pages   = {195--208}
}

@article{DiazTrigo2006,
  author  = {D{\'i}az Trigo, M. and Parmar, A. N. and Boirin, L. and M{\'e}ndez, M. and Kaastra, J. S.},
  title   = {Spectral changes during dipping in low-mass X-ray binaries due to highly-ionized absorbers},
  journal = {A\&A},
  year    = {2006},
  volume  = {445},
  pages   = {179--195}
}

@article{DiazTrigo2009,
  author  = {D{\'i}az Trigo, M. and Parmar, A. N. and Boirin, L. and Motch, C. and Talavera, A. and Balman, {\c S}.},
  title   = {Variations in the dip properties of the low-mass X-ray binary XB 1254-690 observed with XMM-Newton and INTEGRAL},
  journal = {A\&A},
  year    = {2009},
  volume  = {493},
  pages   = {145--157}
}

@ARTICLE{Iaria2007b,
       author = {{Iaria}, R. and {Lavagetto}, G. and {D'A{\'\i}}, A. and {di Salvo}, T. and {Robba}, N.~R.},
        title = "{Chandra observation of the Big Dipper X 1624-490}",
      journal = {A\&A},
         year = 2007,
        month = feb,
       volume = {463},
       number = {1},
        pages = {289-295},
          doi = {10.1051/0004-6361:20065862},
archivePrefix = {arXiv},
       eprint = {astro-ph/0612269},
 primaryClass = {astro-ph},
       adsurl = {https://ui.adsabs.harvard.edu/abs/2007A&A...463..289I}
}

@ARTICLE{Ponti2012,
       author = {{Ponti}, G. and {Fender}, R.~P. and {Begelman}, M.~C. and {Dunn}, R.~J.~H. and {Neilsen}, J. and {Coriat}, M.},
        title = "{Ubiquitous equatorial accretion disc winds in black hole soft states}",
      journal = {MNRAS},
         year = 2012,
        month = may,
       volume = {422},
       number = {1},
        pages = {L11-L15},
          doi = {10.1111/j.1745-3933.2012.01224.x},
archivePrefix = {arXiv},
       eprint = {1201.4172},
 primaryClass = {astro-ph.HE},
       adsurl = {https://ui.adsabs.harvard.edu/abs/2012MNRAS.422L..11P}
}

@article{Watson1985,
  author  = {Watson, M. G. and Willingale, R. and Hertz, P. and Grindlay, J. E. and Seward, F. D.},
  title   = {The X-ray source X1624-490 - A compact binary with extended periodic dips},
  journal = {ApJ},
  year    = {1985},
  volume  = {298},
  pages   = {316--322}
}

@article{Smale2001,
  author  = {Smale, A. P. and Church, M. J. and Ba{\l}uci{\'n}ska-Church, M.},
  title   = {The Ephemeris and Dipping Spectral Behavior of 4U 1624-49},
  journal = {ApJ},
  year    = {2001},
  volume  = {550},
  pages   = {962--972}
}

@ARTICLE{Xiang_07,
       author = {{Xiang}, Jingen and {Lee}, Julia C. and {Nowak}, Michael A.},
        title = "{Using the X-Ray Dust Scattering Halo of 4U 1624-490 to Determine Distance and Dust Distributions}",
      journal = {ApJ},
         year = 2007,
        month = may,
       volume = {660},
       number = {2},
        pages = {1309-1318},
          doi = {10.1086/513308},
archivePrefix = {arXiv},
       eprint = {astro-ph/0701865},
 primaryClass = {astro-ph},
       adsurl = {https://ui.adsabs.harvard.edu/abs/2007ApJ...660.1309X}
}

@ARTICLE{Xiang_09,
       author = {{Xiang}, Jingen and {Lee}, Julia C. and {Nowak}, Michael A. and {Wilms}, J{\"o}rn and {Schulz}, Norbert S.},
        title = "{The Accretion Disk Corona and Disk Atmosphere of 4U 1624-490  as Viewed by the Chandra-High Energy Transmission Grating Spectrometer}",
      journal = {ApJ},
         year = 2009,
        month = aug,
       volume = {701},
       number = {2},
        pages = {984-993},
          doi = {10.1088/0004-637X/701/2/984},
archivePrefix = {arXiv},
       eprint = {0905.3925},
 primaryClass = {astro-ph.HE},
       adsurl = {https://ui.adsabs.harvard.edu/abs/2009ApJ...701..984X}
}

@ARTICLE{Parmar_02,
       author = {{Parmar}, A.~N. and {Oosterbroek}, T. and {Boirin}, L. and {Lumb}, D.},
        title = "{Discovery of narrow X-ray absorption features from the dipping low-mass X-ray binary X 1624-490 with XMM-Newton}",
      journal = {A\&A},
         year = 2002,
        month = may,
       volume = {386},
        pages = {910-915},
          doi = {10.1051/0004-6361:20020281},
archivePrefix = {arXiv},
       eprint = {astro-ph/0202452},
 primaryClass = {astro-ph},
       adsurl = {https://ui.adsabs.harvard.edu/abs/2002A&A...386..910P}
}

@ARTICLE{Caruso_26,
       author = {{Caruso}, E. and {Costantini}, E. and {Degenaar}, N. and {D{\'\i}az Trigo}, M.},
        title = "{An XMM-Newton long look at the accretion disk plasma in the dipping neutron star LMXB 4U 1624─490}",
      journal = {A\&A},
         year = 2026,
        month = jan,
       volume = {705},
          eid = {A176},
        pages = {A176},
          doi = {10.1051/0004-6361/202556100},
archivePrefix = {arXiv},
       eprint = {2510.19177},
 primaryClass = {astro-ph.HE},
       adsurl = {https://ui.adsabs.harvard.edu/abs/2026A&A...705A.176C}
}

@article{Saade2024,
  author = {Saade, M. Lynne and Kaaret, Philip and Gnarini, Andrea and Poutanen, Juri and Ursini, Francesco and Bianchi, Stefano and Bobrikova, Anna and La Monaca, Fabio and Di Marco, Alessandro and Capitanio, Fiamma and et al.},
  title = {X-Ray Polarimetry of the Dipping Accreting Neutron Star 4U 1624--49},
  journal = {ApJ},
  volume = {963},
  pages = {133},
  year = {2024},
  doi = {10.3847/1538-4357/ad235a}
}

@ARTICLE{Gnarini_24,
       author = {{Gnarini}, Andrea and {Lynne Saade}, M. and {Ursini}, Francesco and {Bianchi}, Stefano and {Capitanio}, Fiamma and {Kaaret}, Philip and {Matt}, Giorgio and {Poutanen}, Juri and {Zhang}, Wenda},
        title = "{Constraining the geometry of the dipping atoll 4U 1624{\textendash}49 with X-ray spectroscopy and polarimetry}",
      journal = {A\&A},
         year = 2024,
        month = oct,
       volume = {690},
          eid = {A230},
        pages = {A230},
          doi = {10.1051/0004-6361/202450716},
archivePrefix = {arXiv},
       eprint = {2408.02309},
 primaryClass = {astro-ph.HE},
       adsurl = {https://ui.adsabs.harvard.edu/abs/2024A&A...690A.230G}
}

@ARTICLE{Nitindala_25,
       author = {{Nitindala}, Anagha P. and {Veledina}, Alexandra and {Poutanen}, Juri},
        title = "{X-ray polarization from accretion disk winds}",
      journal = {A\&A},
         year = 2025,
        month = feb,
       volume = {694},
          eid = {A230},
        pages = {A230},
          doi = {10.1051/0004-6361/202453188},
archivePrefix = {arXiv},
       eprint = {2411.18299},
 primaryClass = {astro-ph.HE},
       adsurl = {https://ui.adsabs.harvard.edu/abs/2025A&A...694A.230N}
}

@ARTICLE{Diaz2016,
       author = {{D{\'\i}az Trigo}, M. and {Boirin}, L.},
        title = "{Accretion disc atmospheres and winds in low-mass X-ray binaries}",
      journal = {Astron. Nachr.},
         year = 2016,
        month = may,
       volume = {337},
       number = {4-5},
        pages = {368},
          doi = {10.1002/asna.201612315},
archivePrefix = {arXiv},
       eprint = {1510.03576},
 primaryClass = {astro-ph.HE},
       adsurl = {https://ui.adsabs.harvard.edu/abs/2016AN....337..368D}
}

@article{Begelman1983,
  author  = {Begelman, Mitchell C. and McKee, Christopher F. and Shields, Gregory A.},
  title   = {Compton Heated Winds and Coronae above Accretion Disks. I. Dynamics},
  journal = {ApJ},
  year    = {1983},
  volume  = {271},
  pages   = {70--88},
  doi     = {10.1086/161178}
}

@article{Woods1996,
  author  = {Woods, D. T. and Klein, R. I. and Castor, J. I. and McKee, C. F. and Bell, J. B.},
  title   = {X-Ray--heated Coronae and Winds from Accretion Disks: Time-dependent Two-dimensional Hydrodynamics with Adaptive Mesh Refinement},
  journal = {ApJ},
  year    = {1996},
  volume  = {461},
  pages   = {767--782}
}

@article{Done2018,
  author  = {Done, C. and Tomaru, R. and Takahashi, T.},
  title   = {Thermal winds in stellar mass black hole and neutron star binary systems},
  journal = {MNRAS},
  year    = {2018},
  volume  = {473},
  pages   = {838--848}
}

@article{Tashiro2025,
  author  = {Tashiro, M. and Kelley, R. and Watanabe, S. and others},
  title   = {X-Ray Imaging and Spectroscopy Mission},
  journal = {PASJ},
  year    = {2025},
  volume  = {77},
  pages   = {S1},
  doi     = {10.1093/pasj/psaf023}
}

@article{Ishisaki2025,
  author  = {Ishisaki, Y. and others},
  title   = {Performance and in-orbit calibration of the Resolve instrument onboard XRISM},
  journal = {PASJ},
  year    = {2025}
}

@ARTICLE{DiazTrigo2026,
       author = {{D{\'\i}az Trigo}, M. and {Caruso}, E. and {Costantini}, E. and {Dotani}, T. and {Kohmura}, T. and {Shidatsu}, M. and {Tsujimoto}, M. and {Yoneyama}, T. and {Neilsen}, J. and {Yaqoob}, T. and {Miller}, J.~M.},
        title = "{A highly ionised outflow in the X-ray binary 4U 1624─49 detected with XRISM}",
      journal = {\aap},
         year = 2026,
        month = apr,
       volume = {708},
          eid = {A130},
        pages = {A130},
          doi = {10.1051/0004-6361/202558352},
archivePrefix = {arXiv},
       eprint = {2601.19480},
 primaryClass = {astro-ph.HE},
       adsurl = {https://ui.adsabs.harvard.edu/abs/2026A&A...708A.130D}
}

@ARTICLE{Liao_15,
       author = {{Liao}, Nai-Hui and {Chou}, Yi and {Hsieh}, Hung-En and {Chuang}, Po-Sheng},
        title = "{The Updated Orbital Ephemeris of Dipping Low Mass X-ray Binary 4u 1624-49}",
      journal = {PKAS},
         year = 2015,
        month = sep,
       volume = {30},
       number = {2},
        pages = {593-594},
          doi = {10.5303/PKAS.2015.30.2.593},
       adsurl = {https://ui.adsabs.harvard.edu/abs/2015PKAS...30..593L}
}

@ARTICLE{Ishisaki_25,
       author = {{Ishisaki}, Yoshitaka and {Kelley}, Richard L. and {Awaki}, Hisamitsu and {Balleza}, Jesus C. and {Barnstable}, Kim R. and {Bialas}, Thomas G. and {Boissay-Malaquin}, Rozenn and {Brown}, Gregory V. and {Canavan}, Edgar R. and {Cumbee}, Renata S. and {Carnahan}, Timothy M. and {Chiao}, Meng P. and {Comber}, Brian J. and {Costantini}, Elisa and {den Herder}, Jan-Willem and {Dercksen}, Johannes and {de Vries}, Cor P. and {DiPirro}, Michael J. and {Eckart}, Megan E. and {Ezoe}, Yuichiro and {Ferrigno}, Carlo and {Fujimoto}, Ryuichi and {Gorter}, Nathalie and {Graham}, Steven M. and {Grim}, Martin and {Hartz}, Leslie S. and {Hayakawa}, Ryota and {Hayashi}, Takayuki and {Hell}, Natalie and {Hoshino}, Akio and {Ichinohe}, Yuto and {Ishida}, Manabu and {Ishikawa}, Kumi and {James}, Bryan L. and {Kenyon}, Steven J. and {Kilbourne}, Caroline A. and {Kimball}, Mark O. and {Kitamoto}, Shunji and {Leutenegger}, Maurice A. and {Maeda}, Yoshitomo and {McCammon}, Dan and {Miko}, Joseph J. and {Mizumoto}, Misaki and {Noda}, Hirofumi and {Okajima}, Takashi and {Okamoto}, Atsushi and {Paltani}, Stephane and {Porter}, Frederick S. and {Sato}, Kosuke and {Sato}, Toshiki and {Sawada}, Makoto and {Shinozaki}, Keisuke and {Shipman}, Russell and {Shirron}, Peter J. and {Sneiderman}, Gary A. and {Soong}, Yang and {Szymkiewicz}, Richard and {Szymkowiak}, Andrew E. and {Takei}, Yoh and {Tamura}, Keisuke and {Tsujimoto}, Masahiro and {Uchida}, Yuusuke and {Wasserzug}, Stephen and {Witthoeft}, Michael C. and {Wolfs}, Rob and {Yamada}, Shinya and {Yasuda}, Susumu},
        title = "{Resolve instrument onboard XRISM: design, integration, and instrument test results}",
      journal = {JATIS},
         year = 2025,
        month = oct,
       volume = {11},
          eid = {042023},
        pages = {042023},
          doi = {10.1117/1.JATIS.11.4.042023},
       adsurl = {https://ui.adsabs.harvard.edu/abs/2025JATIS..11d2023I}
}

@article{Harrison2013,
  author = {Harrison, F. A. and Craig, W. W. and Christensen, F. E. and others},
  title = {The Nuclear Spectroscopic Telescope Array (NuSTAR) High-Energy X-Ray Mission},
  journal = {ApJ},
  volume = {770},
  number = {2},
  pages = {103},
  year = {2013},
  doi = {10.1088/0004-637X/770/2/103}
}

@ARTICLE{Kaastra_16,
       author = {{Kaastra}, J.~S. and {Bleeker}, J.~A.~M.},
        title = "{Optimal binning of X-ray spectra and response matrix design}",
      journal = {A\&A},
         year = 2016,
        month = mar,
       volume = {587},
          eid = {A151},
        pages = {A151},
          doi = {10.1051/0004-6361/201527395},
archivePrefix = {arXiv},
       eprint = {1601.05309},
 primaryClass = {astro-ph.IM},
       adsurl = {https://ui.adsabs.harvard.edu/abs/2016A&A...587A.151K}
}

@ARTICLE{Wilms00,
       author = {{Wilms}, J. and {Allen}, A. and {McCray}, R.},
        title = "{On the Absorption of X-Rays in the Interstellar Medium}",
      journal = {ApJ},
         year = 2000,
        month = oct,
       volume = {542},
       number = {2},
        pages = {914-924},
          doi = {10.1086/317016},
archivePrefix = {arXiv},
       eprint = {astro-ph/0008425},
 primaryClass = {astro-ph},
       adsurl = {https://ui.adsabs.harvard.edu/abs/2000ApJ...542..914W}
}

@ARTICLE{Verner_96,
       author = {{Verner}, D.~A. and {Ferland}, G.~J. and {Korista}, K.~T. and {Yakovlev}, D.~G.},
        title = "{Atomic Data for Astrophysics. II. New Analytic FITS for Photoionization Cross Sections of Atoms and Ions}",
      journal = {ApJ},
         year = 1996,
        month = jul,
       volume = {465},
        pages = {487},
          doi = {10.1086/177435},
archivePrefix = {arXiv},
       eprint = {astro-ph/9601009},
 primaryClass = {astro-ph},
       adsurl = {https://ui.adsabs.harvard.edu/abs/1996ApJ...465..487V}
}

@ARTICLE{2020MNRAS.492.5234Z,
       author = {{Zdziarski}, Andrzej A. and {Szanecki}, Micha{\l} and {Poutanen}, Juri and {Gierli{\'n}ski}, Marek and {Biernacki}, Pawe{\l}},
        title = "{Spectral and temporal properties of Compton scattering by mildly relativistic thermal electrons}",
      journal = {MNRAS},
         year = 2020,
        month = mar,
       volume = {492},
       number = {4},
        pages = {5234-5246},
          doi = {10.1093/mnras/staa159},
archivePrefix = {arXiv},
       eprint = {1910.04535},
 primaryClass = {astro-ph.HE},
       adsurl = {https://ui.adsabs.harvard.edu/abs/2020MNRAS.492.5234Z}
}

@ARTICLE{kallman04,
       author = {{Kallman}, T.~R. and {Palmeri}, P. and {Bautista}, M.~A. and {Mendoza}, C. and {Krolik}, J.~H.},
        title = "{Photoionization Modeling and the K Lines of Iron}",
      journal = {ApJS},
         year = 2004,
        month = dec,
       volume = {155},
       number = {2},
        pages = {675-701},
          doi = {10.1086/424039},
archivePrefix = {arXiv},
       eprint = {astro-ph/0405210},
 primaryClass = {astro-ph},
       adsurl = {https://ui.adsabs.harvard.edu/abs/2004ApJS..155..675K}
}

@article{Garcia2014,
  author = {García, J. and Dauser, T. and Lohfink, A. and et al.},
  title = {Improved Reflection Models of Black Hole Accretion Disks: Treating the Angular Distribution of X-Rays},
  journal = {ApJ},
  year = {2014},
  volume = {782},
  pages = {76}
}

@article{Dauser2014,
  author = {Dauser, T. and García, J. and Parker, M. L. and Fabian, A. C. and Wilms, J.},
  title = {Relativistic reflection models: A complete framework for X-ray spectral modeling},
  journal = {MNRAS},
  year = {2014},
  volume = {444},
  pages = {L100}
}

@article{Eker2018,
  author  = {Eker, Z. and Bak{\i}\c{s}, V. and Bilir, S. and Soydugan, F. and Steer, I. and Soydugan, E. and Bak{\i}\c{s}, H. and Ali{\c{c}}avu{\c{s}}, F.},
  title   = {Interrelated main-sequence mass-luminosity, mass-radius, and mass-effective temperature relations},
  journal = {MNRAS},
  volume  = {479},
  pages   = {5491--5511},
  year    = {2018},
  doi     = {10.1093/mnras/sty1834}
}

@article{Pecaut2013,
  author  = {Pecaut, Mark J. and Mamajek, Eric E.},
  title   = {Intrinsic Colors, Temperatures, and Bolometric Corrections of Pre-main-sequence Stars},
  journal = {ApJS},
  volume  = {208},
  number  = {1},
  pages   = {9},
  year    = {2013},
  doi     = {10.1088/0067-0049/208/1/9}
}

@article{Wachter2005,
  author  = {Wachter, Stefanie and Wellhouse, J. W. and Patel, S. K. and Smale, A. P. and Bouchet, P. and Angelini, L.},
  title   = {Near-Infrared and X-Ray Observations of the Dipping Low-Mass X-Ray Binary X1624-490},
  journal = {ApJ},
  volume  = {621},
  number  = {1},
  pages   = {393--399},
  year    = {2005},
  doi     = {10.1086/427420}
}

@article{Guver2009,
  author  = {G{\"u}ver, Tolga and {\"O}zel, Feryal},
  title   = {The relation between optical extinction and hydrogen column density in the Galaxy},
  journal = {MNRAS},
  volume  = {400},
  pages   = {2050--2053},
  year    = {2009},
  doi     = {10.1111/j.1365-2966.2009.15598.x}
}

@article{Cardelli1989,
  author  = {Cardelli, Jason A. and Clayton, Geoffrey C. and Mathis, John S.},
  title   = {The Relationship between Infrared, Optical, and Ultraviolet Extinction},
  journal = {ApJ},
  volume  = {345},
  pages   = {245--256},
  year    = {1989},
  doi     = {10.1086/167900}
}

@inbook{Tauris2006,
  author = {Tauris, T. M. and van den Heuvel, E. P. J.},
  title = {Formation and Evolution of Compact Stellar X-ray Sources},
  booktitle = {Compact Stellar X-ray Sources},
  editor = {Lewin, W. H. G. and van der Klis, M.},
  publisher = {Cambridge University Press},
  year = {2006},
  pages = {623-665}
}

@article{Podsiadlowski2002,
  author = {Podsiadlowski, Ph. and Rappaport, S. and Pfahl, E. D.},
  title = {Evolutionary sequences for low- and intermediate-mass X-ray binaries},
  journal = {ApJ},
  year = {2002},
  volume = {565},
  pages = {1107-1133}
}

@article{ODoherty_2023,
  author = {O'Doherty, T. N. and Bahramian, A. and Miller-Jones, J. C. A. and Goodwin, A. J. and Mandel, I. and Willcox, R. and Atri, P. and Strader, J.},
  title = {An observationally derived kick distribution for neutron stars in binaries},
  journal = {MNRAS},
  volume = {521},
  number = {2},
  pages = {2504--2524},
  year = {2023},
  doi = {10.1093/mnras/stad680}
}

@article{Soberman1997,
  author = {Soberman, G. E. and Phinney, E. S. and van den Heuvel, E. P. J.},
  title = {Stability criteria for mass transfer in binary stellar evolution},
  journal = {A\&A},
  year = {1997},
  volume = {327},
  pages = {620-635}
}

@article{XuLi2007,
  author  = {Xu, Xiao-Jie and Li, Xiang-Dong},
  title   = {Thermal timescale mass transfer rates in intermediate-mass X-ray binaries},
  journal = {A\&A},
  year    = {2007},
  volume  = {476},
  pages   = {1283--1287}
}

@ARTICLE{Ge2024,
       author = {{Ge}, Hongwei and {Tout}, Christopher A. and {Chen}, Xuefei and {Wang}, Song and {Xiong}, Jianping and {Zhang}, Lifu and {Li}, Zhenwei and {Liu}, Qingzhong and {Han}, Zhanwen},
        title = "{Adiabatic Mass Loss in Binary Stars. V. Effects of Metallicity and Nonconservative Mass Transfer{\textemdash}Application in High Mass X-Ray Binaries}",
      journal = {ApJ},
         year = 2024,
        month = nov,
       volume = {975},
       number = {2},
          eid = {254},
        pages = {254},
          doi = {10.3847/1538-4357/ad7ea6},
archivePrefix = {arXiv},
       eprint = {2408.16350},
 primaryClass = {astro-ph.SR},
       adsurl = {https://ui.adsabs.harvard.edu/abs/2024ApJ...975..254G}
}

@article{Belkus2003,
  author  = {Belkus, H. and Van Bever, J. and Vanbeveren, D. and van Rensbergen, W.},
  title   = {The effects of binaries on the evolution of UV spectral features in massive starbursts},
  journal = {A\&A},
  year    = {2003},
  volume  = {400},
  pages   = {429--447},
  doi     = {10.1051/0004-6361:20021814}
}

@article{DiSalvo2009,
  author = {Di Salvo, T. and D'A\`i, A. and Iaria, R. and Burderi, L. and Dov\v{c}iak, M. and Karas, V. and Matt, G. and Papitto, A. and Piraino, S. and Riggio, A. and Robba, N. R. and Santangelo, A.},
  title = {A relativistically smeared spectrum in the neutron star X-ray binary 4U 1705$-$44: looking at the inner accretion disc with X-ray spectroscopy},
  journal = {MNRAS},
  volume = {398},
  number = {4},
  pages = {2022--2027},
  year = {2009},
  doi = {10.1111/j.1365-2966.2009.15306.x}
}

@ARTICLE{Bobrikova_2024a,
       author = {{Bobrikova}, Anna and {Forsblom}, Sofia V. and {Di Marco}, Alessandro and {La Monaca}, Fabio and {Poutanen}, Juri and {Ng}, Mason and {Ravi}, Swati and {Loktev}, Vladislav and {Kajava}, Jari J.~E. and {Ursini}, Francesco and et al.},
        title = "{Discovery of a strong rotation of the X-ray polarization angle in the galactic burster GX 13+1}",
      journal = {A\&A},
         year = 2024,
        month = aug,
       volume = {688},
          eid = {A170},
        pages = {A170},
          doi = {10.1051/0004-6361/202449318},
archivePrefix = {arXiv},
       eprint = {2401.13058},
 primaryClass = {astro-ph.HE},
       adsurl = {https://ui.adsabs.harvard.edu/abs/2024A&A...688A.170B}
}

@ARTICLE{Bobrikova_2024b,
       author = {{Bobrikova}, Anna and {Di Marco}, Alessandro and {La Monaca}, Fabio and {Poutanen}, Juri and {Forsblom}, Sofia V. and {Loktev}, Vladislav},
        title = "{New polarimetric study of the galactic X-ray burster GX 13+1}",
      journal = {A\&A},
         year = 2024,
        month = aug,
       volume = {688},
          eid = {A217},
        pages = {A217},
          doi = {10.1051/0004-6361/202450207},
archivePrefix = {arXiv},
       eprint = {2404.01859},
 primaryClass = {astro-ph.HE},
       adsurl = {https://ui.adsabs.harvard.edu/abs/2024A&A...688A.217B}
}

\begin{appendix}
 
\section{Atomic transitions used in the Fe--K analysis}
\label{app:lines}

\begin{table}[th]
\caption{Identified absorption lines: rest-frame energies and atomic transitions.}
\label{tab:lines_id_add}
\centering
\scriptsize
\setlength{\tabcolsep}{3pt}

\begin{tabular}{lll}
\hline\hline
$E_{0}$ (keV) & Line ID & Atomic transition \\
\hline
6.6366 & \ion{Fe}{xxv} He$\alpha$ (z) &
$1s^2\,{}^{1}S_{0} \rightarrow 1s\,2s\,{}^{3}S_{1}$ \\
6.6676 & \ion{Fe}{xxv} He$\alpha$ (y) &
$1s^2\,{}^{1}S_{0} \rightarrow 1s\,2p\,{}^{3}P_{1}$ \\
6.7004 & \ion{Fe}{xxv} He$\alpha$ (w) &
$1s^2\,{}^{1}S_{0} \rightarrow 1s\,2p\,{}^{1}P_{1}$ \\
6.9521 & \ion{Fe}{xxvi} Ly$\alpha_{2}$ &
$1s\,{}^{2}S_{1/2} \rightarrow 2p\,{}^{2}P_{1/2}$ \\
6.9732 & \ion{Fe}{xxvi} Ly$\alpha_{1}$ &
$1s\,{}^{2}S_{1/2} \rightarrow 2p\,{}^{2}P_{3/2}$ \\
7.7316 & \ion{Ni}{xxvii} He$\alpha$ (z) &
$1s^2\,{}^{1}S_{0} \rightarrow 1s\,2s\,{}^{3}S_{1}$ \\
7.8056 & \ion{Ni}{xxvii} He$\alpha$ (w) &
$1s^2\,{}^{1}S_{0} \rightarrow 1s\,2p\,{}^{1}P_{1}$ \\
7.8811 & \ion{Fe}{xxv} He$\beta$ (w3) &
$1s^2\,{}^{1}S_{0} \rightarrow 1s\,3p\,{}^{1}P_{1}$ \\
8.2464 & \ion{Fe}{xxvi} Ly$\beta_{2}$ &
$1s\,{}^{2}S_{1/2} \rightarrow 3p\,{}^{2}P_{1/2}$ \\
8.2530 & \ion{Fe}{xxvi} Ly$\beta_{1}$ &
$1s\,{}^{2}S_{1/2} \rightarrow 3p\,{}^{2}P_{3/2}$ \\
8.2956 & \ion{Fe}{xxv} He$\gamma$ (w4) &
$1s^2\,{}^{1}S_{0} \rightarrow 1s\,4p\,{}^{1}P_{1}$ \\
8.4875 & \ion{Fe}{xxv} He$\delta$ (w5) &
$1s^2\,{}^{1}S_{0} \rightarrow 1s\,5p\,{}^{1}P_{1}$ \\
\hline\hline
\end{tabular}

\end{table}

\section{XRISM/Resolve and NuSTAR light curves}
\label{app:lightcurves}

\begin{figure}[ht] 
\centering 
\scriptsize

\includegraphics[width=0.8\columnwidth]
{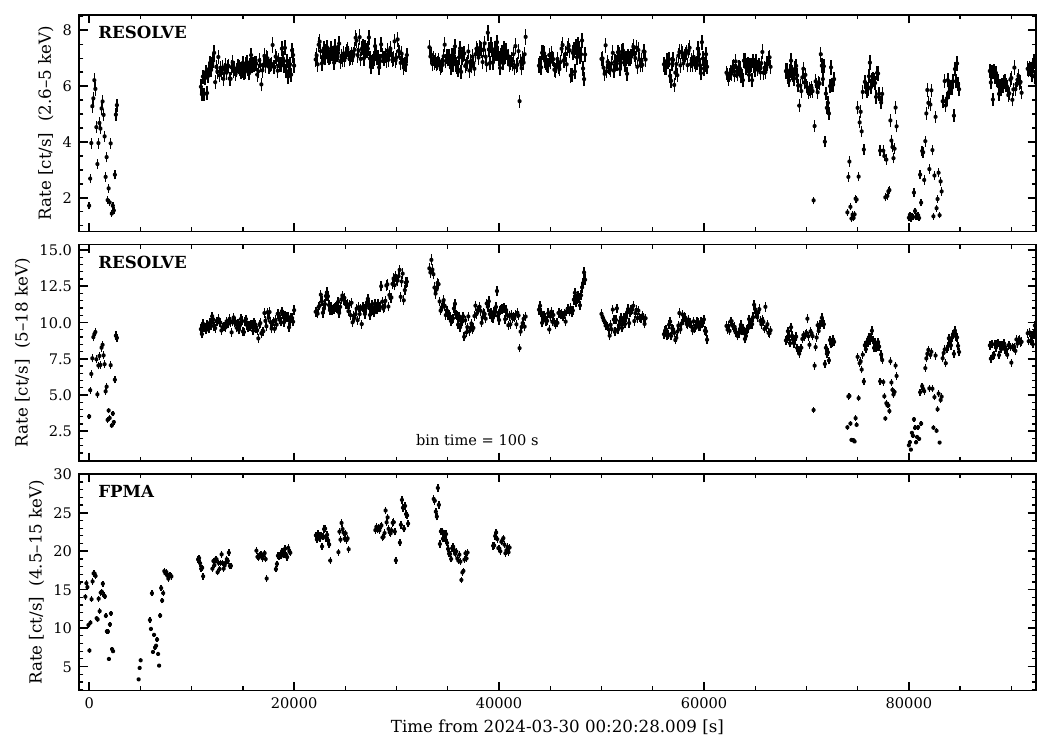}

\vspace{3mm}

\includegraphics[width=0.8\columnwidth]
{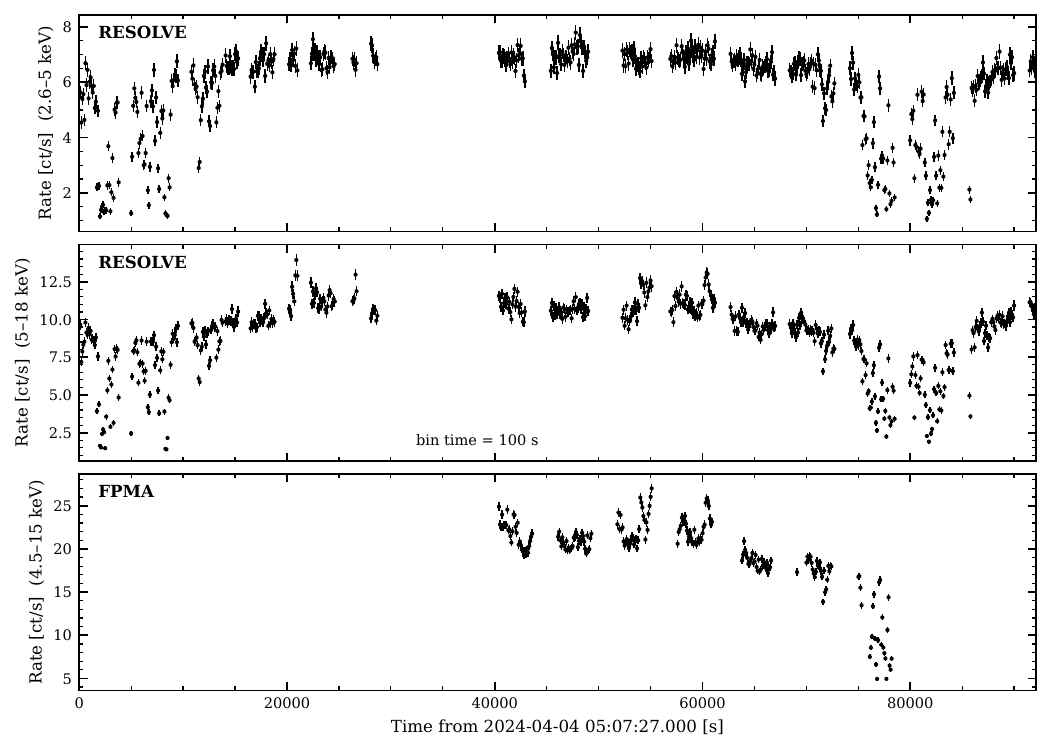}

\caption{Light curves of 4U~1624$-$490 for the two coordinated
XRISM and NuSTAR observations, with Obs.~1 shown in the top
panel and Obs.~2 in the bottom panel. For each observation, the
top and middle panels show the Resolve count rate in the
2.6--5~keV and 5--18~keV bands, respectively, while the bottom
panel shows the simultaneous NuSTAR/FPMA count rate in the
4.5--15~keV band. All light curves are rebinned to 100~s.}
\label{fig:curves}
\end{figure}

 \begin{figure}[ht] 
\centering 
\scriptsize
\includegraphics[width=\columnwidth]
{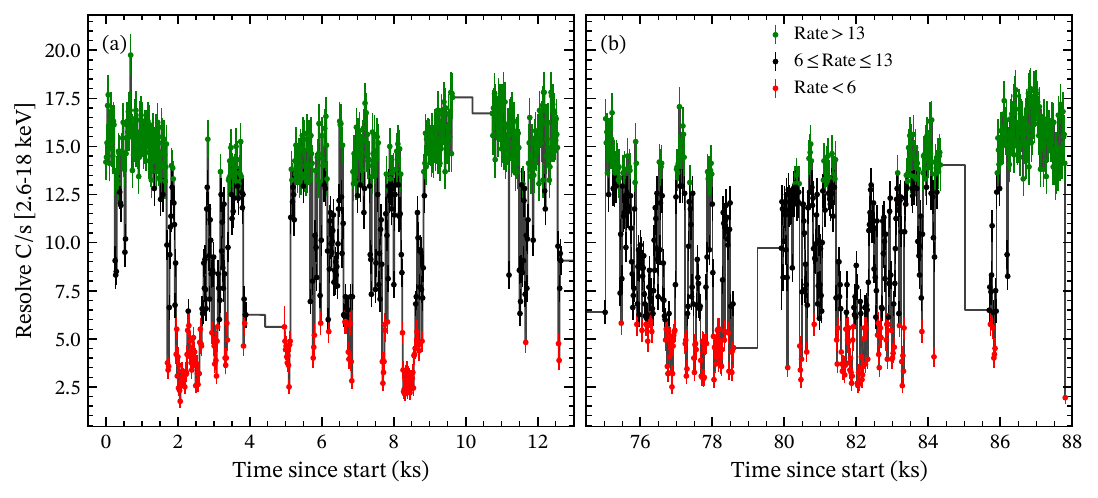}

\caption{Background-subtracted light curve of Observation~2
in the orbital-phase interval 0.95--1.12. The points are
coloured according to count rate: red for ${\rm Rate}<6$,
black for $6\leq{\rm Rate}\leq13$, and green for
${\rm Rate}>13$. The bin time is 16~s.}

\label{fig:dip_lightcurve}
 
\end{figure}

\section{Fe--K line profiles in the remaining orbital-phase intervals and broadband residuals}
\label{app:fek_residuals}

 \begin{figure*}[!h] 
\centering
\includegraphics[width=0.98\textwidth]{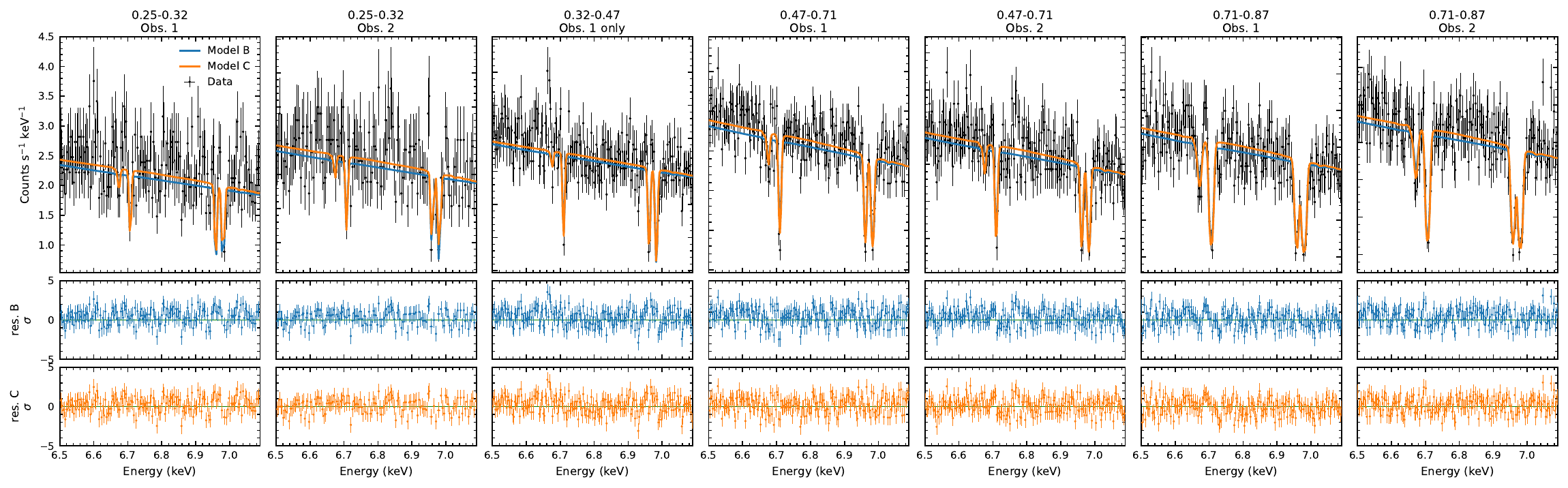}
\captionof{figure}{Same as Fig.~\ref{fig:feK_modelB_modelC_main}, but for the
remaining orbital-phase intervals. For the 0.32--0.47 interval, only the Obs.~1 Resolve spectrum is shown for display purposes, although the spectral fit also includes Obs.~2, fitted over the restricted 2--16 keV range because of its lower statistical quality. The comparison confirms that Model B provides an adequate description of the narrow Fe--K absorption features over the full phase-resolved dataset.}
\label{fig:feK_modelB_modelC_app}
\end{figure*}

 \begin{figure*}[!h] 
    \centering
    \includegraphics[width=0.98\textwidth]{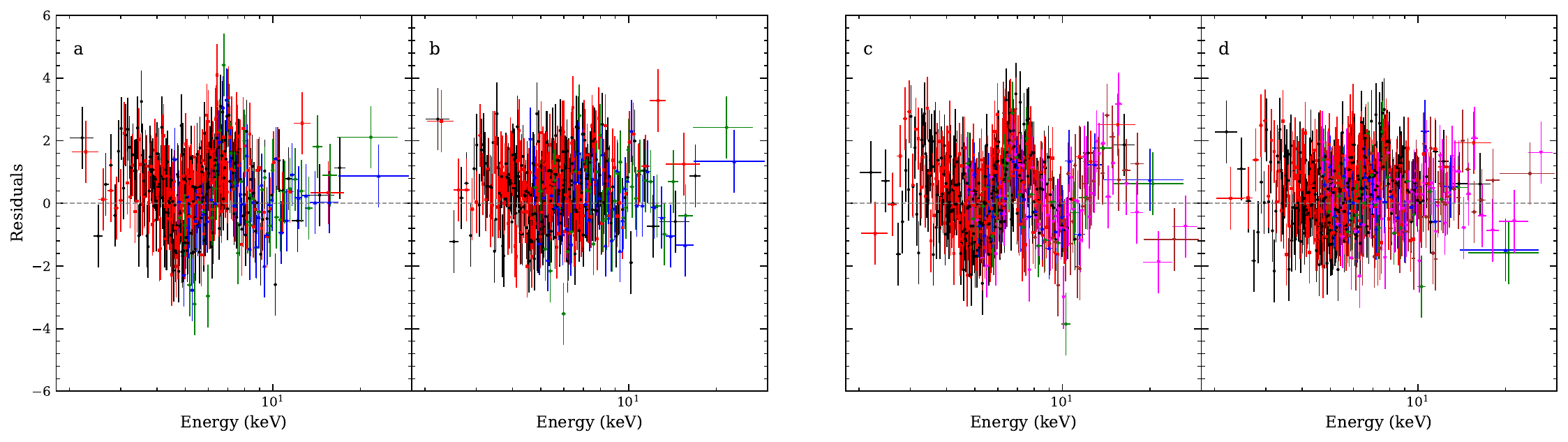}
    \captionof{figure}{Comparison of the residuals, in units of standard deviations, obtained with \texttt{Model B} and \texttt{Model C} for two representative orbital-phase intervals, namely interval~2 ($\phi=0.12$--$0.25$) and interval~5 ($\phi=0.47$--$0.71$). The left and right panels show the residuals obtained with \texttt{Model B} and \texttt{Model C}, respectively. The colours identify the different spectra included in each phase-resolved joint fit: black and red correspond to the two XRISM/Resolve spectra, green and magenta to the NuSTAR/FPMA spectra, and blue and brown to the corresponding NuSTAR/FPMB spectra. The spectra are graphically rebinned for display purposes only.}
    \label{fig:residuals_modelB_modelC}
 \end{figure*}

\clearpage

\section{Best-fit parameter values for Model B}
\label{app:spectral_fits_Model_B_c}

\begin{strip}

\begin{minipage}{\textwidth}
\centering

\captionof{table}{Best--fit parameters obtained with Model B.}
\label{tab:modelB_all}

\begin{threeparttable}

\scriptsize
\setlength{\tabcolsep}{3.2pt}
\renewcommand{\arraystretch}{1.05}

\begin{tabular}{@{}llccccccc@{}}
\toprule

Component & Parameter
& 0.95--1.12
& 0.12--0.25
& 0.25--0.32
& 0.32--0.47
& 0.47--0.71
& 0.71--0.87
& 0.87--0.95 \\

\midrule

\texttt{const} & Resolve
& --
& \makecell{[1]\\$1.011\pm0.006$}
& \makecell{[1]\\$0.972\pm0.009$}
& \makecell{[1]\\$0.96\pm0.02$}
& \makecell{[1]\\$0.980\pm0.005$}
& \makecell{[1]\\$1.011\pm0.006$}
& \makecell{[1]\\$1.029\pm0.009$} \\

\texttt{const} & FPMA
& --
& \makecell{$1.116\pm0.008$\\--}
& \makecell{$1.119\pm0.009$\\--}
& \makecell{$1.087\pm0.006$\\--}
& \makecell{$1.058\pm0.010$\\$1.037\pm0.005$}
& \makecell{--\\$1.079^{+0.007}_{-0.005}$}
& \makecell{--\\$1.091\pm0.009$} \\

\texttt{const} & FPMB
& --
& \makecell{$1.107\pm0.007$\\--}
& \makecell{$1.113\pm0.009$\\--}
& \makecell{$1.080^{+0.008}_{-0.004}$\\--}
& \makecell{$1.044\pm0.010$\\$1.024\pm0.005$}
& \makecell{--\\$1.062\pm0.006$}
& \makecell{--\\$1.070\pm0.009$} \\

\addlinespace[2pt]

\texttt{TBabs} & $N_{\rm H}$
& [4.0]
& [4.0]
& [4.0]
& [4.0]
& [4.0]
& [4.0]
& [4.0] \\

\texttt{TBpcf} & $N_{\rm H,pc}$
& $15.4\pm0.6$
& $11.6^{+0.4}_{-0.2}$
& $10.5^{+0.6}_{-1.0}$
& $12.0^{+1.5}_{-0.5}$
& $13.1\pm0.3$
& $12.4\pm0.6$
& $11.7^{+0.9}_{-0.4}$ \\

& $pcf$
& $0.696^{+0.011}_{-0.015}$
& $0.709^{+0.009}_{-0.012}$
& $0.72^{+0.03}_{-0.02}$
& $0.72\pm0.03$
& $0.661^{+0.007}_{-0.010}$
& $0.673^{+0.010}_{-0.007}$
& $0.723\pm0.013$ \\

\texttt{PartCov} & $f$
& $0.74\pm0.02$
& $0.61^{+0.03}_{-0.02}$
& $>0.99$
& $0.81^{+0.06}_{-0.14}$
& $0.56^{+0.04}_{-0.02}$
& $0.67\pm0.03$
& $0.61\pm0.02$ \\

\addlinespace[2pt]

\texttt{warmabs} & $N_{\rm H,WA}$
& $30^{+4}_{-3}$
& $44^{+6}_{-3}$
& $11^{+12}_{-3}$
& $13^{+12}_{-6}$
& $40^{+6}_{-8}$
& $31^{+4}_{-3}$
& $47^{+8}_{-5}$ \\

& $\log\xi_{\rm WA}$
& $3.22^{+0.01}_{-0.04}$
& $3.493^{+0.016}_{-0.007}$
& $3.84^{+0.07}_{-0.15}$
& $3.78^{+0.09}_{-0.07}$
& $3.845^{+0.008}_{-0.027}$
& $3.65^{+0.04}_{-0.02}$
& $3.38^{+0.02}_{-0.05}$ \\

& $v_{\rm turb}$ (km\,s$^{-1}$)
& $161^{+18}_{-16}$
& $127^{+14}_{-10}$
& [50]
& [50]
& $90^{+28}_{-20}$
& $217\pm17$
& $215^{+14}_{-26}$ \\

& $z$ ($10^{-4}$)
& $-2.5253^{+0.0002}_{-0.0011}$
& $-5.70^{+0.04}_{-0.02}$
& $-11.21^{+0.33}_{-0.15}$
& $-14.6^{+0.8}_{-0.6}$
& $-14.7\pm0.2$
& $-9.50^{+0.06}_{-0.13}$
& $-3.693^{+0.008}_{-0.016}$ \\

\addlinespace[2pt]

\texttt{thcomp} & $\Gamma$
& $3.02^{+0.08}_{-0.05}$
& \makecell{$3.002^{+0.009}_{-0.012}$\\$2.80\pm0.02$}
& \makecell{$3.17^{+0.06}_{-0.03}$\\$2.86^{+0.05}_{-0.04}$}
& \makecell{$2.55^{+0.06}_{-0.03}$\\$2.83^{+0.05}_{-0.07}$}
& \makecell{$3.202^{+0.019}_{-0.014}$\\$2.933^{+0.008}_{-0.014}$}
& \makecell{$3.19\pm0.03$\\$2.931^{+0.006}_{-0.050}$}
& $2.99^{+0.02}_{-0.06}$ \\

& $kT_{\rm e}$ (keV)
& [3.35]
& $3.342^{+0.009}_{-0.012}$
& $3.71^{+0.22}_{-0.12}$
& $3.17^{+0.13}_{-0.06}$
& $3.581^{+0.008}_{-0.019}$
& $3.516\pm0.007$
& $3.35\pm0.10$ \\

\addlinespace[2pt]

\texttt{bbodyrad} & $kT_{\rm bb}$ (keV)
& $1.034^{+0.007}_{-0.008}$
& $1.051\pm0.006$
& $1.139^{+0.011}_{-0.014}$
& $1.019^{+0.012}_{-0.031}$
& $1.098^{+0.002}_{-0.010}$
& $1.072^{+0.005}_{-0.003}$
& $1.046^{+0.005}_{-0.011}$ \\

& Norm$_{\rm bb}$
& $131.3^{+0.9}_{-1.8}$
& $124.6^{+0.3}_{-0.6}$
& $96\pm8$
& $152^{+16}_{-8}$
& $115.2^{+4.4}_{-0.2}$
& $119.5^{+0.3}_{-3.3}$
& $117.9^{+11.0}_{-0.4}$ \\

\addlinespace[2pt]

\texttt{fit} & $\chi^2$ (d.o.f.)
& $2461.76\,(2233)$
& $5304.61\,(4964)$
& $3408.67\,(3392)$
& $2980.86\,(2989)$
& $5471.96\,(5279)$
& $4795.65\,(4756)$
& $3891.94\,(3652)$ \\

& $\Delta\chi^2$
& 2135
& 2120
& 200
& 322
& 876
& 1937
& 1387 \\

\bottomrule
\end{tabular}

\begin{tablenotes}[flushleft]
\footnotesize
\item[]
Notes.
The dip column refers to the high-count-rate spectrum extracted
from the 0.95--1.12 orbital-phase interval.
$N_{\rm H}$, $N_{\rm H,pc}$, and $N_{\rm H,WA}$ are given in units of
$10^{22}\,\mathrm{cm^{-2}}$.
Parameters reported in square brackets were held fixed during
the spectral fitting.
Quoted uncertainties correspond to the 90\% confidence level.
For rows with two entries in a given orbital-phase interval,
the upper and lower values refer to Obs.~1 and Obs.~2, respectively.
$\Delta\chi^2$ values are computed relative to Model A.
\end{tablenotes}

\end{threeparttable}

\end{minipage}

\end{strip}

\section{On the physical meaning of the velocity offset $v_{0}$}
\label{app:v0}
\nolinenumbers
The line-of-sight velocity curve of the ionised absorber was modelled
with the sinusoidal function
\begin{equation}
v(\phi)=v_0+K\sin[2\pi(\phi-\phi_0)],
\end{equation}
where $K$ is the semi-amplitude of the orbital modulation and $v_0$ is
the constant velocity offset. The seven-point fit presented in
Sect.~\ref{sect:radial_velocity}
 gives $v_0=-285\pm9$ km s$^{-1}$.

If $K$ approximately traces the orbital motion of the neutron star, the constant term $v_{0}$ should not be interpreted simply as the intrinsic velocity of the absorbing plasma. Rather, it represents the phase-averaged line-of-sight offset measured in the observer frame and may include several contributions,
\begin{equation}
v_{0}\simeq v_{\rm abs}+\gamma+v_{\rm obs},
\end{equation}
where \(v_0\) is the  line-of-sight velocity offset measured in the observer frame, $v_{\rm abs}$ is the bulk velocity of the absorber relative to the compact object, $\gamma$ is the systemic radial velocity of the binary, and $v_{\rm obs}$ accounts for the motion of the observer.

For the two XRISM observations considered here, the observer-related term \(v_{\rm obs}\) includes both the projected Earth motion along the line of sight, which contributes an apparent blueshift of about \(-23\) km s\(^{-1}\), and the low-Earth-orbit motion of the satellite, which largely averages out over the long exposures. Correcting the measured offset \(v_0 \simeq -285~\mathrm{km\,s^{-1}}\) for this contribution still leaves a residual velocity of   \(v_{\rm abs}+\gamma\simeq-262~\mathrm{km\,s^{-1}}\), showing that the observed blueshift cannot be explained by the observer motion alone.

A further source of uncertainty is the systemic velocity of the binary, \(\gamma\), namely the line-of-sight velocity of the centre of mass of the system with respect to the observer. Following \citet{ODoherty_2023}, we describe this quantity through a prior of the form \(P(\gamma \mid D)=\mathcal{N}(\gamma_G, v_p^2)\), where \(\mathcal{N}(\gamma_G,v_p^2)\) is a Gaussian distribution centred on \(\gamma_G\) with standard deviation \(v_p\). In this expression, \(D\) is the source distance, \(\gamma_G\) is the radial velocity expected from Galactic rotation at the source position and distance, and \(v_p\) represents the characteristic peculiar-velocity dispersion of the relevant X-ray binary population. For the Galactic coordinates of 4U~1624$-$490 and an assumed distance of \(\sim 15\) kpc, we estimate \(\gamma_G \simeq +3\ {\rm km\,s^{-1}}\), while \(v_p \sim 100\ {\rm km\,s^{-1}}\) provides a plausible scale for the residual systemic motion, consistent with the NS-LMXB subsample considered by \citet{ODoherty_2023}, whose inferred velocity distribution has a mode near \(90\ {\rm km\,s^{-1}}\) and includes several systems with peaks around \(100\ {\rm km\,s^{-1}}\).
This prior implies that the systemic velocity of the binary is plausibly of the order of \(\gamma \approx +3 \pm 100~\mathrm{km\,s^{-1}}\).

Adopting an estimate for \(\gamma\), the quantity \(v_{\rm abs}+\gamma\simeq-262~\mathrm{km\,s^{-1}}\) implies a representative absorber-frame velocity of roughly 
\begin{equation}
v_{\rm abs}\approx -265 \pm 100~\mathrm{km\,s^{-1}},
\end{equation}
so that the mean absorber motion remains consistently blueshifted. The exact decomposition of $v_{0}$ is uncertain, but the qualitative conclusion is robust: the ionised plasma is not stationary on average and is most naturally interpreted as showing net outward motion.

\section{Consistency of a low-mass donor interpretation}
\label{app:lowmass_donor}

In Sect.~\ref{sec:donor_implications} we discussed the donor-star properties under the assumption that the observed velocity modulation of the ionised absorber traces, at least to first order, the orbital motion of the neutron star. If this assumption is relaxed, the measured velocity amplitude cannot be directly converted into a binary mass function. In that case, a low-mass donor cannot be excluded on dynamical grounds alone. Here we examine whether such an interpretation is consistent with the orbital period, Roche geometry, and the near-infrared counterpart.

As a representative low-mass case, we consider a donor mass of $M_2=0.5\;M_\odot$, assuming $M_1=1.4\;M_\odot$ and $P_{\rm orb}=0.869896$~d. Kepler's third law gives
$ 
a 
\simeq 3.30\times10^{11}\ {\rm cm}$.
Using the Eggleton approximation, with $q=M_2/M_1$, the Roche-lobe radius of the donor is 
$ 
R_{{\rm L},2}\simeq 9.7\times10^{10}\ {\rm cm}
\simeq 1.4\;R_\odot $.
This radius is substantially larger than that of an unevolved main-sequence star of the same mass, for which $R_2\simeq0.5\;R_\odot$ is expected.  Therefore, a Roche-lobe-filling low-mass donor at the orbital period of 4U~1624--490 would have to be substantially enlarged with respect to a normal main-sequence star, for example because it is evolved, inflated, or stripped.

The near-infrared counterpart provides an additional consistency check. Using $m_{K_{\rm s}}=18.3\pm0.1$ mag, $A_{K_{\rm s}}\simeq2.12$ mag, and a distance modulus $\mu\simeq15.88$ mag, the corresponding absolute magnitude is
$
M_{K_{\rm s}}
\simeq m_{K_{\rm s}}-A_{K_{\rm s}}-\mu
\simeq 0.3 $.
This is significantly brighter than expected for the photosphere of an unevolved $0.5\;M_\odot$ main-sequence star. Therefore, if the donor were of low mass, the observed $K_{\rm s}$-band emission would have to include a substantial contribution from the accretion flow, the outer disc rim, and/or X-ray irradiation. The infrared magnitude alone therefore does not rule out a low-mass donor, but it could not be interpreted unambiguously as the photospheric emission of a normal low-mass companion.  

The thermal timescale also illustrates the non-standard nature of such a solution. For a Roche-lobe-filling donor with $M_2=0.5\;M_\odot$ and $R_2\simeq1.4\;R_\odot$, the Kelvin--Helmholtz time is
$
t_{\rm KH}\simeq
5.5\times10^6
\left({L_2}/{L_\odot}\right)^{-1}
{\rm yr}$.
For $L_2\simeq1\;L_\odot$, this corresponds to a thermal time of several Myr. If the donor were an evolved or stripped object with $L_2\simeq5$--$10\;L_\odot$, the thermal time would decrease to $\sim5\times10^5$--$10^6$ yr, comparable to the thermal timescale inferred for the more massive donor solution discussed in Sect.~\ref{sec:donor_implications}. Thus, the Kelvin--Helmholtz timescale does not exclude a low-mass donor, but it requires the star to be in a non-standard evolutionary state rather than being an unevolved low-mass dwarf.

We conclude that, if the absorber velocity modulation is not associated with the orbital motion of the neutron star, a low-mass donor remains possible. However, a standard unevolved low-mass main-sequence companion is not naturally compatible with the 20.9 hr orbital period, because it would be too compact to fill its Roche lobe and too faint to account for the observed near-infrared counterpart without substantial non-stellar emission. A viable low-mass interpretation would therefore require an evolved, inflated, or stripped donor, together with a significant contribution from the accretion flow and/or irradiation to the observed near-infrared light.

\section{Evolutionary implications of the IMXB interpretation}
\label{app:imxb_evolution}

For a late B-type donor compatible with a B9~V--B9.5~V star, the
relevant evolutionary response is expected to occur on a thermal
timescale. Using the Kelvin--Helmholtz estimate $ 
t_{\rm KH}\simeq {G M_2^2}/{(R_2 L_2)}$,
and adopting the dwarf parameters tabulated by
\citet{Pecaut2013}, namely
$(M_2,R_2,L_2)\simeq(2.75\,M_\odot,2.49\,R_\odot,
72.4\,L_\odot)$ for a B9~V star and
$(2.68\,M_\odot,2.45\,R_\odot,63.1\,L_\odot)$ for a
B9.5~V star, we obtain
$ t_{\rm KH}\simeq(1.3-1.5)\times10^6\ {\rm yr}$.
This is much shorter than the nuclear lifetime of such stars and
supports the interpretation that, if the donor is an intermediate-mass
late B star filling its Roche lobe, mass transfer is likely proceeding
on or near the donor thermal timescale.

The corresponding characteristic donor mass-loss rate is
$ 
\dot{M}_{\rm th}\sim{M_2}/{t_{\rm KH}}
\simeq(1.8-2.1)\times10^{-6}\,
M_\odot\,{\rm yr}^{-1}$.
This estimate should be regarded as an order-of-magnitude indication of the donor mass-loss rate rather than of the rate effectively accreted by the neutron star. It nevertheless shows that, under the IMXB interpretation, the system would naturally evolve in a strongly non-conservative regime.
In this case, mass transfer would not be expected to follow the slow secular evolution typical of ordinary LMXBs. Rather, it would more naturally proceed on the donor thermal-adjustment timescale and would likely be strongly non-conservative \citep{Soberman1997, Tauris2006, Podsiadlowski2002}.  Detailed evolutionary calculations show that IMXBs with neutron-star accretors can experience thermal-timescale mass transfer at rates far above the Eddington accretion rate, while still avoiding immediate dynamical runaway when the donor retains a predominantly radiative envelope \citep{XuLi2007, Ge2024}. This point is relevant here because the preferred solution implies a mass ratio $q\simeq2$ and a donor compatible with a B9~V--B9.5~V   star, namely a regime in which rapid Roche-lobe overflow is physically plausible without necessarily leading at once to common-envelope evolution.

\section{A possible circumbinary contribution to the local neutral absorber}
\label{app:circumbinary_absorber}

As a simple order-of-magnitude estimate, we ask whether a non-conservative mass-transfer rate of order
$
\dot M \sim 10^{-6}\,M_\odot\,{\rm yr^{-1}} \simeq 6.3 \times10^{19}\ {\rm g\,s^{-1}}
$
could sustain the equivalent local neutral hydrogen column inferred from the X-ray spectra, $N_{\rm H}\sim10^{23}\ {\rm cm^{-2}}$, if a fraction of the expelled material accumulates in a circumbinary structure.  

Following the circumbinary-ring formalism introduced by \citet{Soberman1997}, mass lost from the binary can be parametrised as forming a circumbinary structure at radius \(R_{\rm CB}=\eta a\), where \(a\) is the binary separation. In subsequent applications of this prescription, \(\eta\simeq1.3\) is commonly treated as a lower-limit value, while \(\eta\simeq2.25\) is often adopted for the first stable circumbinary ring \citep[e.g.][]{Belkus2003}.
In this formalism, \(R_{\rm CB}\) is measured from the centre of mass of the binary system.

Assuming, for simplicity, that the expelled gas is diluted over a quasi-spherical surface of radius $r$ and moves with a bulk velocity of
$
v \sim 100\ {\rm km\,s^{-1}}
$
in a non-conservative mass-transfer flow, the continuity equation gives
$
\rho(r) \sim {\dot M}/{(4\pi r^{2}v)},
$
and therefore the hydrogen number density can be written as
$ 
n_{\rm H}(r) \sim {\rho}/({\mu_{\rm H} m_{\rm p}})
      \sim {\dot M}/({4\pi r^{2}\mu_{\rm H} m_{\rm p}v}),
$
where $\mu_{\rm H}$ is the mean mass per hydrogen nucleus in units of the proton mass, $m_{\rm p}$ is the proton mass, and for near-solar composition we adopt $\mu_{\rm H}\simeq1.4$.
For the first stable circumbinary configuration we adopt \(R_{\rm CB}=\eta a\) with \(\eta=2.25\). Since the circumbinary structure is centred on the binary centre of mass, the corresponding distance measured from the donor is \(r\simeq R_{\rm CB}-a_2\), where \(a_2=aM_1/(M_1+M_2)\) is the distance of \(M_2\) from the centre of mass. For the adopted binary parameters, this gives \(r\simeq 1.92\,a\).
Assuming \(\dot M\sim10^{-6}\,M_\odot\,{\rm yr^{-1}}\), \(v\sim100\ {\rm km\ s^{-1}}\), and \(\mu_{\rm H}=1.4\), the hydrogen number density at this radius is \(n_{\rm H}\simeq(3.2\pm0.7)\times10^{11}\ {\rm cm^{-3}}\).

We compare this value with the equivalent neutral hydrogen column density inferred for the local absorber in the X-ray spectra, \(N_{\rm H}\sim10^{23}\ {\rm cm^{-2}}\). Under the simplifying assumption of an approximately uniform density along the line of sight through the circumbinary material, this implies an effective path length along the line of sight of \(\Delta s \simeq N_{\rm H}/n_{\rm H}\approx3.2\times10^{11}\ {\rm cm}\simeq0.75\;a\).
This simple order-of-magnitude estimate shows that, if a fraction of the transferred mass is lost from the binary and accumulates in a circumbinary structure, the characteristic size and density of such material are sufficient to account for the observed equivalent neutral hydrogen column density local to the system.

An additional question is whether such a circumbinary absorber would necessarily be transient.
 A simple order-of-magnitude estimate suggests that this is not required. Approximating the absorbing structure as a torus, its volume can be written as
$
V \sim 2\pi^{2} R H^{2},
$
where \(R\) is the major radius and \(H\) is the minor radius.
Adopting \(R\simeq R_{\rm CB}\), and using the characteristic density \(n_{\rm H}\) inferred from the continuity equation together with the effective path length \(\Delta s\), the vertical thickness of the ring is related to the inclination through
$
z \simeq \Delta s \cos i,
$
so that, in the toroidal approximation,
$
H \simeq z/2.
$
The total gas mass contained in the structure is then
$ 
M_{\rm torus}\sim \mu_{\rm H}m_{\rm p}\,n_{\rm H}\,V
\sim \mu_{\rm H}m_{\rm p}\,n_{\rm H}\,(2\pi^{2}RH^{2}),
$ 
and the corresponding replenishment timescale is
$
t_{\rm refill}\sim {M_{\rm torus}}/{\dot M} \sim3.8 \times 10^{-11} M_{\odot} /(10^{-6} M_{\odot}/yr)\sim 20\ {\rm min}.
$
Since the inferred torus mass is very small, the required absorbing material can, in principle, be replenished on a timescale much shorter than the duration of the non-conservative mass-transfer phase itself. This suggests that, once formed, such a circumbinary structure need not be purely transient, but could instead be maintained quasi-persistently through continuous replenishment by the ongoing outflow.

\section{Absorber location in the Compton-heated corona/wind framework}
\label{app:compton_wind}

To place the ionised absorber in a physical context, we compare the inferred absorber location with the characteristic scales of a Compton-heated atmosphere/wind. In the framework of \citet{Begelman1983}, the relevant length scale is the Compton radius,
$R_{\rm C}={G M_1 \mu m_{\rm p}}/{(k_{\rm B} T_{\rm C})}$,
where \(M_1\) is the compact-object mass (assumed 1.4 $M_{\odot}$), \(\mu\simeq0.6\) is the mean molecular weight, \(m_{\rm p}\) is the proton mass, \(k_{\rm B}\) is the Boltzmann constant, and \(T_{\rm C}\) is the Compton temperature of the irradiating spectrum. Physically, \(R_{\rm C}\) is the radius at which gas heated to \(T_{\rm C}\) has a thermal speed comparable to the local escape velocity. Gas located well inside \(R_{\rm C}\) is generally expected to remain gravitationally bound, forming an X-ray heated corona, whereas thermal expansion and wind launching become increasingly favoured at sufficiently large radii.

A second key quantity is the critical luminosity \(L_{\rm crit}\), namely the luminosity above which Compton heating is efficient enough to drive a thermal wind. Following \citet{Begelman1983}, this can be written as
$
L_{\rm crit} \simeq 0.03\,T_{{\rm C},8}^{-1/2}\,L_{\rm Edd},
$
where \(T_{{\rm C},8}=T_{\rm C}/10^{8}\,{\rm K}\) and \(L_{\rm Edd}\) is the Eddington luminosity. For the value adopted in this work, \(T_{\rm C}=10^{7}\) K, this gives \(L_{\rm crit}\simeq 0.095\,L_{\rm Edd}\) and $R_{\rm C}\simeq 1.35\times10^{11} \,{\rm cm}$.

The values reported in Table~\ref{tab:nfe_tot} show that in all orbital-phase intervals the source is above the critical luminosity, with \(L/L_{\rm crit}\simeq 2.3\)--3.1. Therefore, from an energetic point of view, the system lies in a regime in which Compton heating can, in principle, sustain a thermal outflow. At the same time, the inferred absorber locations span \(r/R_{\rm C}\simeq 0.32\)--1.1, but all the quoted \(r\) values are upper limits. The corresponding \(r/R_{\rm C}\) ratios must therefore also be treated as upper limits: the data are compatible with gas located near the regime where thermal launching becomes plausible, but they do not require the absorber to reside in a freely escaping large-scale wind, since it may still lie at smaller radii within a bound or only marginally unbound Compton-heated atmosphere.

The larger upper limits inferred for the near-dip and dip intervals should
be interpreted with caution. They may reflect increased complexity or
stratification of the absorbing structure along the line of sight at these
phases, rather than a direct geometrical displacement of a single
homogeneous absorber to larger radii. The corresponding line-of-sight
velocities are therefore the most susceptible to systematic uncertainties
related to unresolved multi-phase absorption.

\begin{table}
\caption{Luminosity ratios and inferred absorber location for the different orbital-phase intervals.}
\label{tab:nfe_tot}
\centering
\scriptsize

\setlength{\tabcolsep}{4pt}
\renewcommand{\arraystretch}{1.0}

\begin{tabular}{lcccc}
\toprule
Interval
& $L/L_{\rm Edd}$
& $L/L_{\rm crit}$
& $r$ ($10^{10}\,\mathrm{cm}$)
& $r/R_{\rm C}$ \\
\midrule

0.12--0.25
& $0.24\pm0.01$
& $2.50\pm0.12$
& $8.0\pm1.0$
& $0.59\pm0.07$ \\

0.25--0.32
& $0.25\pm0.01$
& $2.62\pm0.13$
& $4.5\pm2.2$
& $0.34\pm0.16$ \\

0.32--0.47
& $0.29\pm0.01$
& $3.04\pm0.15$
& $5.9\pm2.6$
& $0.43\pm0.19$ \\

0.47--0.71
& $0.26\pm0.01$
& $2.70\pm0.14$
& $5.0\pm1.1$
& $0.37\pm0.08$ \\

0.71--0.87
& $0.24\pm0.01$
& $2.52\pm0.13$
& $5.3\pm0.7$
& $0.39\pm0.05$ \\

0.87--0.95
& $0.23\pm0.01$
& $2.39\pm0.12$
& $7.1\pm1.2$
& $0.53\pm0.09$ \\

0.95--1.12, CR$>13$
& $0.23\pm0.01$
& $2.46\pm0.12$
& $13.3\pm0.2$
& $0.99\pm0.17$ \\

\bottomrule
\end{tabular}

\tablefoot{
The quoted radial distances are estimated from the observed ionisation
parameter and column density under the assumption
$\Delta r/r \leq 1$.
The luminosity ratios were derived assuming a source distance of
15~kpc, a neutron-star mass of $1.4\,M_\odot$, and a 5\% uncertainty
on the flux, while no uncertainty on the distance was included.
The absorber distance was estimated from
$r=L/(\xi N_{\rm H,WA})$, and the values reported in the table
therefore correspond to upper limits obtained under the assumption
$\Delta r/r=1$.
The Compton radius was computed for $M_1=1.4\,M_\odot$ and
$T_{\rm C}=10^{7}$~K.
}

\end{table}

The absorber distance was estimated as $r = {L}/{(\xi N_{\rm H,WA})}$,
assuming \(\Delta r/r=1\), so the values listed in Table~\ref{tab:nfe_tot} remain indicative rather than definitive. This caveat is especially important for the intervals with \(r/R_{\rm C}\sim 0.3\)--0.6, where the plasma could still belong to a bound or marginally bound heated atmosphere rather than to a fully developed outflow. The measured blueshift is therefore more robustly interpreted as evidence for outward motion than as proof that the gas is already part of a freely expanding large-scale wind.

As an order-of-magnitude estimate, the characteristic launch radius can be obtained by equating the observed velocity to the local escape speed,
$ R_{\rm launch}\sim {2GM_1}/{v^2}.$
For \(M_1=1.4\,M_\odot\) and \(v\simeq170\)--370 km s\(^{-1}\), this yields
$
R_{\rm launch}\sim 2.7\times10^{11} - 1.3\times10^{12}\ {\rm cm}.
$
Assuming an outer disc radius \(R_{\rm disk}\simeq0.9R_{\rm L1}\simeq1.25\times10^{11}\) cm, these values correspond to
$
R_{\rm launch}\sim 2.2 - 10.5\,R_{\rm disk}.
$

Taken at face value, these estimates would place the launching site well beyond the nominal outer disc radius, which is unlikely if the outflow originates in the accretion disc. This comparison, therefore, indicates that the measured blueshift should not be identified directly with the full dynamical velocity of the flow. Rather, the observed line-of-sight velocity most likely underestimates the true outflow speed, either because of projection effects or because the absorbing plasma is detected close to the wind base, before reaching its terminal velocity. This does not, however, exclude the possibility that the phase dependence of the line centroid retains a dynamical component linked to the orbital motion of the compact object. Instead, it suggests that the absolute velocity offset and the phase-modulated component need not encode the same physical information. In this sense, the measured blueshift is more naturally interpreted as evidence for outward motion in a dense wind base or failed-wind configuration than as a direct tracer of the true launch radius.
The current data provide evidence for outward motion, but not for unbound motion.

\section{Physical interpretation of the reflection component}
\label{app:reflection}

Reflection is statistically required in all phase intervals, including
the high-count-rate dip spectrum. By accounting for the broad Fe--K and
broadband curvature, it permits a cleaner separation between reflected
emission and ionised absorption and should therefore be regarded as an
intrinsic component of the spectral decomposition rather than as a
phenomenological correction.

Outside the dipping interval, the reflector is generally found to be moderately ionised, with $\log \xi_{\rm refl}\simeq 2.75$ whenever this parameter is well constrained. This is consistent with ionisation levels reported for disc reflection in neutron-star low-mass X-ray binaries, for example in 4U~1705$-$44, where $\log \xi \sim 2.7$ was found by \cite{DiSalvo2009}.
 Such values imply that the reflected emission arises in optically thick material exposed to a strong irradiating flux, naturally consistent with the accretion-disc surface. In this regime, the broad excess observed around 6--7 keV is more naturally interpreted as part of an ionised reflection continuum than as an isolated broad emission line.

The \texttt{relxillNS} model provides an adequate description of
the reflection component across all orbital phases.  For each phase-resolved spectrum, the inner disc radius $R_{\rm in}$ was explored in preliminary fits and then fixed to the corresponding best-fit value in the final fits, in order to reduce degeneracies among the reflection parameters. 
 This does not demonstrate that the innermost disc is absent, but it indicates that the present spectra do not require the observed reflection to be dominated by the most strongly relativistic disc regions. The reflected emission can therefore be explained without invoking a major contribution from only a few gravitational radii, suggesting either that the observed reflection is weighted toward larger disc radii or that any inner-disc contribution is partly diluted by the complex absorption and high-inclination geometry.

 This interpretation is broadly consistent with recent studies of
4U 1624--490, although the polarimetric evidence should not be
regarded as a unique diagnostic of disc reflection. \citet{Saade2024}   reported a significant non-dip polarisation degree and argued
that the polarised signal is naturally associated with Comptonised
emission plus an additional contribution from photons reflected by
the disc. Their analysis further suggested that the observed
polarisation is difficult to explain with the immediate neutron-star
vicinity alone, and is instead consistent with a more extended
Comptonising geometry, such as a slab-like corona above the disc.
\citet{Gnarini_24} likewise found that the broadband spectrum
requires a reflection component from the accretion disc and showed
that, during dips, the highly ionised absorber becomes denser and
less ionised.

However, the interpretation of X-ray polarisation in high-inclination
neutron-star LMXBs is not unique. Scattering in an extended
equatorial disc wind may also contribute to the observed
polarisation, and recent radiative-transfer calculations have shown
that accretion-disc winds can produce polarisation degrees comparable
to those observed in highly inclined X-ray binaries \citep{Nitindala_25,Bobrikova_2024a,Bobrikova_2024b}. Therefore, in the
present work we do not use the polarisation properties as an
independent proof of reflection. The reflection component is instead
motivated by the broadband XRISM+NuSTAR spectral residuals, while
polarimetry provides a complementary indication that
reprocessed/scattered radiation is important in the system. A
dedicated phase-resolved spectro-polarimetric analysis would be
required to separate quantitatively the wind-scattering/re-emission
and reflection contributions.

\end{appendix}

\end{document}